\documentclass[useAMS,usenatbib]{mn2e}
\pdfoutput=1
\usepackage{amsmath}
\usepackage{amsfonts}
\usepackage{bm}
\usepackage{booktabs}
\usepackage{natbib}
\usepackage{graphicx}
\usepackage{wasysym}
\usepackage[colorlinks=true,backref=page,bookmarksdepth=2]{hyperref}

\newcommand{\Log}{ \operatorname{Log}}

\allowdisplaybreaks

\title[The Stark problem in the Weierstrassian formalism]
{
The Stark problem in the Weierstrassian formalism
}

\author[Francesco Biscani and Dario Izzo]{Francesco Biscani$^1$\thanks{Previously
at the Advanced Concepts Team, ESA. E-mail: bluescarni@gmail.com} and Dario Izzo$^2$\\
$^{1}$Castle Mews 27, St. Thomas Street,\\
$\hphantom{^{2}}$Oxford, Oxfordshire, OX1 1JR, United Kingdom\\
$^{2}$ESA -- Advanced Concepts Team, European Space Research Technology Centre (ESTEC),\\
$\hphantom{^{2}}$Keplerlaan 1, Postbus 299, 2200 AG Noordwijk The Netherlands
}

\begin{document}

\date{\today}

\pagerange{\pageref{firstpage}--\pageref{lastpage}}

\maketitle

\label{firstpage}

\begin{abstract}
We present a new general, complete closed-form solution of the three-dimensional Stark problem in terms of Weierstrass elliptic and
related functions. With respect
to previous treatments of the problem, our analysis is exact and valid for all values of the external force field, and it is expressed
via unique formul\ae{} valid for all initial conditions and parameters of the system. The simple form of the solution allows us to perform a thorough
investigation of the properties of the dynamical system, including the identification of quasi-periodic and periodic orbits, the formulation
of a simple analytical criterion to determine the boundness of the trajectory, and the characterisation of the equilibrium points.
\end{abstract}

\section{Introduction}
The dynamical system consisting of a test particle subject simultaneously to an inverse-square central field and to a force field
constant both in magnitude and direction is known under multiple denominations. Historically, this system was first studied
in detail in the context of particle physics (where it is known as \emph{Stark problem} \citep{stark_beobachtungen_1914}) in connection with
the shifting and splitting of spectral lines of atoms and molecules in the presence of an external static electric field.

In astrophysics and dynamical astronomy, the Stark problem is sometimes known as the \emph{accelerated Kepler problem}, and it is studied
in several contexts. Models based on the accelerated Kepler problem have been used to study the excitation of planetary orbits
by stellar jets in protoplanetary disks and to explain the origin of the eccentricities of extrasolar planets
\citep{namouni_origin_2005,namouni_accelerated_2007,namouni_excitation_2013}. The Stark problem
has also been used in the study of the dynamics of dust grains in the Solar System \citep{belyaev_dynamics_2010,pastor_influence_2012}.

In astrodynamics, the Stark problem is relevant in connection to the \emph{continuous-thrust problem}, describing the dynamics of spacecrafts
equipped with ion thrusters. In such a context, the trajectory of the spacecraft is often considered as a series of non-Keplerian arcs
resulting from the simultaneous action of the gravitational acceleration and the constant thrust provided by the engine \citep{sims_preliminary_1999}.

From a purely mathematical perspective, the importance of the Stark problem lies mainly in fact that it belongs to the very restrictive class
of Liouville-integrable dynamical systems of classical mechanics \citep{arnold_mathematical_1989}.
Action-angle variables for the Stark problem can be introduced in a perturbative fashion, as explained in
\citet{born_mechanics_1927} and \citet{berglund_averaged_2001}.

Different types of solutions to the Stark problem are available in the literature. If the constant acceleration field is much smaller than
the Keplerian attraction along the orbit of the test particle, the problem can be treated in a perturbative fashion,
and the (approximate) solution is expressed as the variation in time of the Keplerian (or Delaunay) orbital elements of the osculating
orbit \citep{vinti_effects_1966,berglund_averaged_2001,namouni_accelerated_2007,belyaev_dynamics_2010,pastor_influence_2012}. A different
approach is based on regularisation procedures such as the Levi-Civita and Kustaanheimo-Stiefel transformations
\citep{kustaanheimo_perturbation_1965,saha_interpreting_2009}, which yield exact solutions in a set of variables related to the cartesian ones
through a rather complex nonlinear transformation
\citep{kirchgraber_problem_1971,rufer_trajectory_1976,poleshchikov_one_2004}. A third way exploits the formulation of the Stark problem in parabolic coordinates
to yield an exact solution in terms of Jacobi elliptic functions and integrals \citep{lantoine_complete_2011}.

The aim of this paper is to introduce and examine a new solution to the Stark problem that employs the Weierstrassian elliptic and related functions.
The main features of our solution can be summarised as follows:
\begin{itemize}
 \item it is an exact (i.e., non-perturbative), closed-form and explicit solution;
 \item it is expressed as a set of unique formul\ae{} independent of the values of the initial conditions and of the parameters of the system;
 \item it is a solution to the full three-dimensional Stark problem (whereas many previous solutions deal only with the restricted case in which
 the motion is confined to a plane).
\end{itemize}
The simple form of our solution allows us to examine thoroughly the dynamical features of the Stark problem, and to derive several new results (e.g., regarding
questions of (quasi) periodicity and boundness of motion). Our method of solution is in some sense close to the one employed in \citet{lantoine_complete_2011}.
However, we believe that our solution offers several distinct advantages:
\begin{itemize}
 \item by adopting the Weierstrassian formalism (instead of the Jacobian one), we sidestep the issue of categorising the solutions based on the nature of the roots of the polynomials generating
 the differential equations, and thus our formul\ae{} do not depend on the initial conditions or on the parameters of the system;
 \item we provide explicit formul\ae{} for the three-dimensional case;
 \item we avoid introducing a second time transformation in the solution.
\end{itemize}
These advantages are critical in providing new insights in the dynamics of the Stark problem. On the other hand, the use of the Weierstrassian formalism presents a few additional
difficulties with respect to the approach described in \citet{lantoine_complete_2011}, the most notable of which is probably the necessity of operating in the complex domain. Throughout the paper,
we will highlight these difficulties and address them from the point of view of the actual implementation of the formul\ae{} describing our solution to the Stark problem.

In this paper, we will focus our attention specifically on the full three-dimensional Stark problem,
where the motion of the test particle is not confined to a plane,
and we will only hint occasionally at the bidimensional case (where instead the motion is constrained to a plane).
\section{Problem formulation}
From a dynamical point of view, the Stark problem is equivalent to a one-body gravitational problem with an additional force field
which is constant both in magnitude and direction. The corresponding Lagrangian in cartesian coordinates
$\boldsymbol{r} = \left(x,y,z\right)$
and velocities $\boldsymbol{v} = \left(\dot{x},\dot{y},\dot{z}\right)$ is then
\begin{equation}
L\left(\boldsymbol{v};\boldsymbol{r}\right)=\frac{1}{2}v^{2}+\frac{\mu}{r}+\varepsilon z,
\label{eq:orig_lagr}
\end{equation}
where the inertial coordinate system has been centred on the central body,
$v = \left| \boldsymbol{v}\right|$, $r = \left| \boldsymbol{r}\right|$, $\mu$ is the gravitational parameter
of the system and $\varepsilon > 0$ is the constant acceleration imparted to the test particle by the force field.
Without loss of generality, the coordinate system has been oriented so that the force field is directed towards the positive $z$ axis.

Following the lead of \citet{epstein_zur_1916} and \citet{born_mechanics_1927}, we proceed
by expressing the Lagrangian in parabolic coordinates $\left( \xi, \eta, \phi \right)$ via the coordinate transformation
\begin{align}
x & =\xi\eta\cos\phi, & \dot{x} & =\left(\dot{\xi}\eta+\xi\dot{\eta}\right)\cos\phi-\xi\eta\dot{\phi}\sin\phi,\label{eq:x_para}\\
y & =\xi\eta\sin\phi, & \dot{y} & =\left(\dot{\xi}\eta+\xi\dot{\eta}\right)\sin\phi+\xi\eta\dot{\phi}\cos\phi,\\
z & =\frac{\xi^{2}-\eta^{2}}{2}, & \dot{z} & =\xi\dot{\xi}-\eta\dot{\eta},\label{eq:z_para}
\end{align}
where $\xi \geq 0$, $\eta \geq 0$ and $\phi \in \left( -\pi,\pi\right]$ is the azimuthal angle.
The inverse transformation from cartesian to parabolic coordinates is
\begin{align}
\xi & =\sqrt{r+z}, & \dot{\xi} & =\frac{\dot{r}+\dot{z}}{2\sqrt{r+z}},\label{eq:xi_inv}\\
\eta & =\sqrt{r-z}, & \dot{\eta} & =\frac{\dot{r}-\dot{z}}{2\sqrt{r-z}},\label{eq:eta_inv}\\
\phi & =\arctan\left(y,x\right), & \dot{\phi} & =\frac{\dot{y}x-\dot{x}y}{x^{2}+y^{2}}\label{eq:phi_inv},
\end{align}
where $\dot{r}=\left(\boldsymbol{v}\cdot\boldsymbol{r}\right)/r$ and $\arctan$ is the two-argument inverse tangent function.
In the new coordinate system,
\begin{align}
v^{2} & =\left(\xi^{2}+\eta^{2}\right)\left(\dot{\xi}^{2}+\dot{\eta}^{2}\right)+\xi^{2}\eta^{2}\dot{\phi}^{2},\\
r & =\frac{\xi^{2}+\eta^{2}}{2},
\end{align}
and the Lagrangian becomes
\begin{multline}
L=\frac{1}{2}\left[\left(\xi^{2}+\eta^{2}\right)\left(\dot{\xi}^{2}+\dot{\eta}^{2}\right)+\xi^{2}\eta^{2}\dot{\phi}^{2}\right]\\
+\frac{2\mu}{\xi^{2}+\eta^{2}}+\varepsilon\frac{\xi^{2}-\eta^{2}}{2}.
\end{multline}
Switching now to the Hamiltonian formulation through a Legendre
transformation, the momenta are defined as
\begin{align}
p_{\xi} & =\frac{\partial L}{\partial\dot{\xi}}=\left(\xi^{2}+\eta^{2}\right)\dot{\xi},\label{eq:p_xi}\\
p_{\eta} & =\frac{\partial L}{\partial\dot{\eta}}=\left(\xi^{2}+\eta^{2}\right)\dot{\eta},\label{eq:p_eta}\\
p_{\phi} & =\frac{\partial L}{\partial\dot{\phi}}=\xi^{2}\eta^{2}\dot{\phi},\label{eq:p_phi}
\end{align}
and the Hamiltonian is written as
\begin{align}
\mathcal{H} & =\dot{\xi}p_{\xi}+\dot{\eta}p_{\eta}+\dot{\phi}p_{\phi}-L\\
 & =\frac{1}{2}\frac{p_{\xi}^{2}+p_{\eta}^{2}}{\xi^{2}+\eta^{2}}+\frac{1}{2}\frac{p_{\phi}^{2}}{\xi^{2}\eta^{2}}-\frac{2\mu}{\xi^{2}+\eta^{2}}-\varepsilon\frac{\xi^{2}-\eta^{2}}{2}.
\label{eq:Ham_def}
\end{align}
Since the coordinate $\phi$ is absent from the Hamiltonian, the momentum $p_{\phi}$ is a constant of motion. It can
be checked by substitution that $p_{\phi}$ is the $z$ component of
the total angular momentum of the system. Thus, when $p_{\phi}$ vanishes,
the motion is confined to a plane perpendicular to the
$xy$ plane and intersecting the origin, and we can refer to this subcase as the
\emph{bidimensional} problem (as opposed to the \emph{three-dimensional}
problem when $p_{\phi}$ is not null).

We now employ a Sundman regularisation \citep{sundman_memoire_1912}, introducing the fictitious time $\tau$ via the differential relation
\begin{equation}
dt=\left(\xi^{2}+\eta^{2}\right)d\tau,\label{eq:fic_time}
\end{equation}
and the new, identically null, function
\begin{equation}
\mathcal{H}_{\tau}\left(p_\xi,p_\eta,p_\phi;\xi,\eta,\phi\right)=\left(\mathcal{H}-h\right)\left(\xi^{2}+\eta^{2}\right),
\end{equation}
where $h$ is the energy constant of the system (obtained
by substituting the initial conditions into the expression of $\mathcal{H}$).
We have then for $p_{\xi}$ and $\xi$
\begin{align}
\frac{dp_{\xi}}{d\tau} & =\frac{dp_{\xi}}{dt}\frac{dt}{d\tau}=-\frac{\partial\mathcal{H}}{\partial\xi}\left(\xi^{2}+\eta^{2}\right)=-\frac{\partial\mathcal{H}_{\tau}}{\partial\xi},\\
\frac{d\xi}{d\tau} & =\frac{d\xi}{dt}\frac{dt}{d\tau}=\frac{\partial\mathcal{H}}{\partial p_{\xi}}\left(\xi^{2}+\eta^{2}\right)=\frac{\partial\mathcal{H}_{\tau}}{\partial p_{\xi}},
\end{align}
and, similarly for $p_{\eta}$, $\eta$, $p_{\phi}$ and $\phi$,
\begin{align}
\frac{dp_{\eta}}{d\tau} & = -\frac{\partial\mathcal{H}_{\tau}}{\partial\eta},\\
\frac{d\eta}{d\tau} & = \frac{\partial\mathcal{H}_{\tau}}{\partial p_{\eta}},\\
\frac{dp_{\phi}}{d\tau} & = -\frac{\partial\mathcal{H}_{\tau}}{\partial\phi},\\
\frac{d\phi}{d\tau} & = \frac{\partial\mathcal{H}_{\tau}}{\partial p_{\phi}}.
\end{align}
$\mathcal{H}_\tau$ can thus be considered as an Hamiltonian function
describing the evolution of the system in fictitious time\footnote{
This regularisation procedure is sometimes referred to as \emph{Poincar\'{e} trick} or \emph{Poincar\'{e} time transform}
\citep{siegel_lectures_1971,carinena_time_1988,saha_interpreting_2009}.}. Explicitly,
\begin{multline}
\mathcal{H}_{\tau}=-\varepsilon\frac{\xi^{4}}{2}-h\xi^{2}+\frac{1}{2}p_{\xi}^{2}+\frac{1}{2}\frac{p_{\phi}^{2}}{\xi^{2}}\\
+\varepsilon\frac{\eta^{4}}{2}-h\eta^{2}+\frac{1}{2}p_{\eta}^{2}+\frac{1}{2}\frac{p_{\phi}^{2}}{\eta^{2}}-2\mu,
\end{multline}
and the Hamiltonian $\mathcal{H}_\tau$ has thus been separated into the two independent
constants of motion
\begin{align}
\alpha_{1} & =-\varepsilon\frac{\xi^{4}}{2}-h\xi^{2}+\frac{1}{2}p_{\xi}^{2}+\frac{1}{2}\frac{p_{\phi}^{2}}{\xi^{2}},\label{eq:alpha1}\\
\alpha_{2} & =\varepsilon\frac{\eta^{4}}{2}-h\eta^{2}+\frac{1}{2}p_{\eta}^{2}+\frac{1}{2}\frac{p_{\phi}^{2}}{\eta^{2}}.\label{eq:alpha2}
\end{align}
These constants represent the conservation of a component of the generalised Runge-Lenz vector \citep{redmond_generalization_1964}.
By inversion of $\alpha_{1}$ and $\alpha_{2}$ for $p_{\xi}$ and
$p_{\eta}$, Hamilton's equations finally yield
\begin{align}
p_\xi=\frac{d\xi}{d\tau}&=\pm\frac{1}{\xi}\sqrt{\varepsilon\xi^{6}+2h\xi^{4}+2\alpha_{1}\xi^{2}-p_{\phi}^{2}},\label{eq:xi_diffeq}\\
p_\eta=\frac{d\eta}{d\tau}&=\pm\frac{1}{\eta}\sqrt{-\varepsilon\eta^{6}+2h\eta^{4}+2\alpha_{2}\eta^{2}-p_{\phi}^{2}},\label{eq:eta_diffeq}\\
\frac{d\phi}{d\tau}&=p_\phi\left(\frac{1}{\xi^2}+\frac{1}{\eta^2}\right).\label{eq:phi_diffeq}
\end{align}
The solution of the Stark problem has thus been reduced to the integration by quadrature
of eqs. \eqref{eq:xi_diffeq}--\eqref{eq:phi_diffeq}.
Before proceeding, it is useful to outline the general features of the
functions on the right-hand side of eqs. \eqref{eq:xi_diffeq} and
\eqref{eq:eta_diffeq}.

\subsection{Study of $p_{\xi}\left(\xi\right)$ and $p_{\eta}\left(\eta\right)$}
\label{subsec:poly_study}
Both $p_{\xi}\left(\xi\right)$ and $p_{\eta}\left(\eta\right)$ are
functions of $\xi$ and $\eta$ symmetric with respect to both the
horizontal and vertical axes. The zeroes of both functions are given
by the roots of the bicubic polynomial radicands on the right-hand
side of eqs. \eqref{eq:xi_diffeq} and \eqref{eq:eta_diffeq}. Hence,
the number of real roots of $p_{\xi}\left(\xi\right)$ and $p_{\eta}\left(\eta\right)$
will depend on the initial conditions and on the physical parameters
of the system (namely, the gravitational parameter and the value
of the constant force field).

\begin{figure*}
\includegraphics[width=1\textwidth]{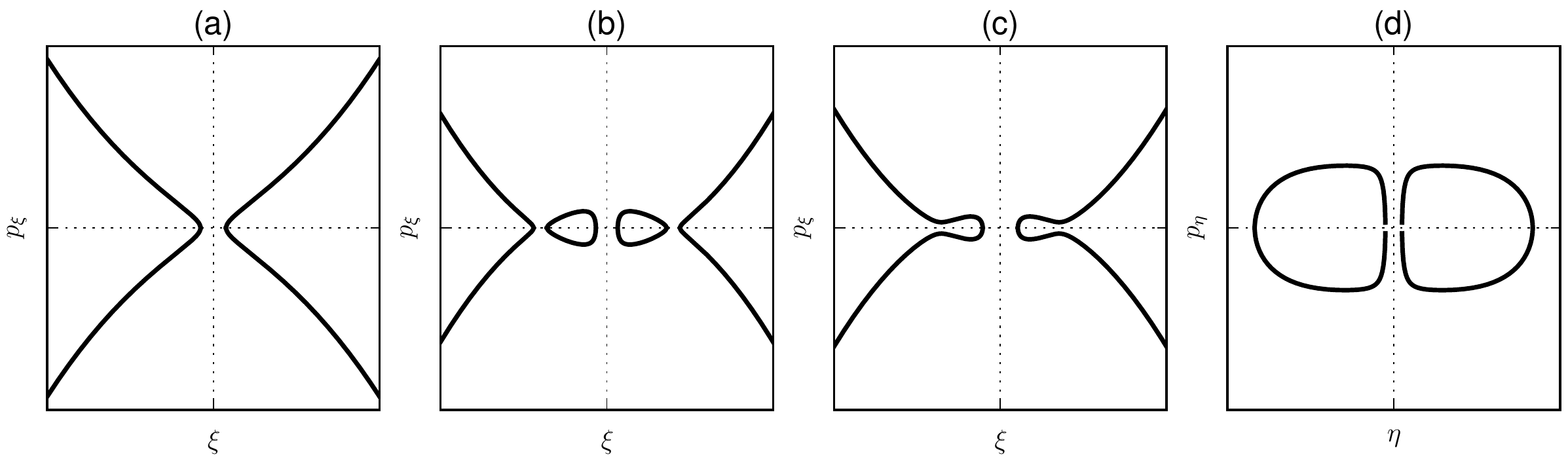}

\caption{Representative phase plots in the
three-dimensional case. The evolution of $\xi$ and $p_{\xi}$ (a,b,c) can be bound
or unbound, depending on the initial conditions and on the values
of the parameters of the system. By contrast, the evolution of $\eta$ and $p_{\eta}$
is always bound (d).\label{fig:tridimensional_case}}
\end{figure*}

For any given set of initial conditions, it is clear that 
the polynomial radicand on the right-hand side of eq. \eqref{eq:xi_diffeq}
will tend to $+\infty$ for $\xi\to\pm\infty$, since $\varepsilon>0$ by definition.
Thus, $p_{\xi}\left(\xi\right)$
will always tend to $\pm\infty$ in the limit $\xi\to\pm\infty$.
Conversely, for $\eta\to\pm\infty$,
the radicand in $p_{\eta}\left(\eta\right)$
will eventually start assuming negative values, thus implying the
existence of a real root. For both $p_{\xi}\left(\xi\right)$ and
$p_{\eta}\left(\eta\right)$, moving along the horizontal axis towards
the origin from the initial conditions means encountering another root,
as for $\xi=\eta=0$ both functions result in the computation of the
square root of the negative value $-p_{\phi}^{2}$. This also implies
that, in the three-dimensional problem, the trajectories in the phase planes $\left(\xi,p_\xi \right)$ and $\left(\eta,p_\eta\right)$
will not cross the vertical axes, and $p_{\xi}\left(\xi\right)$ and $p_{\eta}\left(\eta\right)$
always have at least two real roots. Figure \ref{fig:tridimensional_case}
shows a selection of representative trajectories in the phase space for
$p_{\xi}\left(\xi\right)$ and $p_{\eta}\left(\eta\right)$ in the three-dimensional case.

\begin{figure*}
\includegraphics[width=1\textwidth]{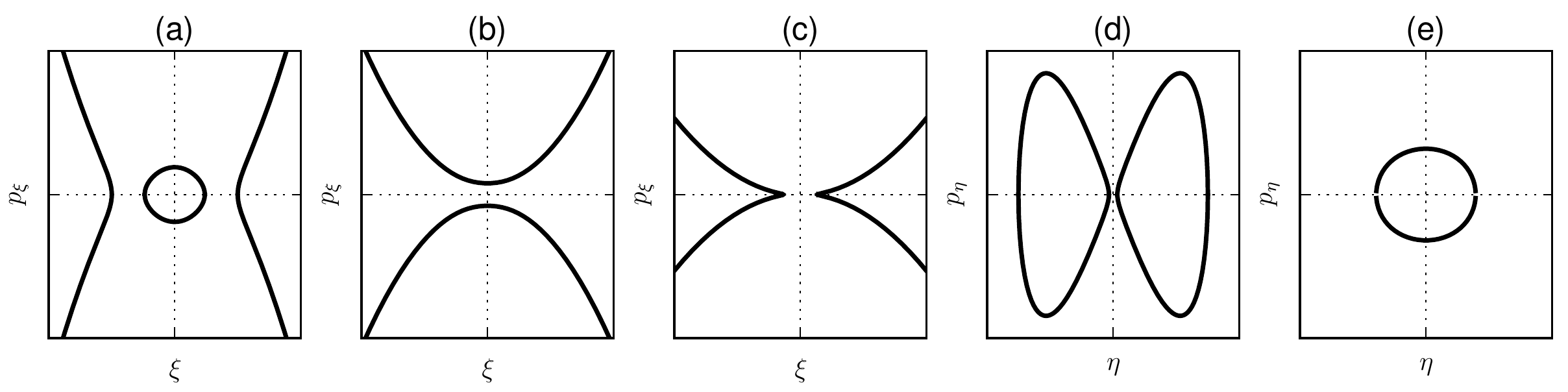}

\caption{Representative phase plots in the
bidimensional case. Similarly to the three-dimensional case, the evolution
of $\xi$ and $p_{\xi}$ (a,b,c) can be bound or unbound, whereas the evolution
of $\eta$ and $p_{\eta}$ is always bound (d,e).\label{fig:bidimensional_case}}
\end{figure*}

The bidimensional case requires a separate analysis. When $p_{\phi}$ is null,
the bicubic polynomials collapse to biquadratic polynomials (via the
inclusion of the external factors $1/\xi$ and $1/\eta$). As in the
three-dimensional case, the evolution of $p_{\xi}$ can be either bound
or unbound, while the evolution of $p_{\eta}$ is always bound.
The first difference is that, when $\alpha_{1}>0$, $p_{\xi}$ might
have no real roots. Secondly, when the signs of the constants $\alpha_{1}$ and $\alpha_{2}$
are positive, $p_{\xi}$ and $p_{\eta}$ assume real values for $\xi=0$
and $\eta=0$, and the trajectories in the phase plane thus seemingly cross the vertical axes.
Physically, the conditions $\xi=0$ and $\eta=0$ correspond (via eqs. \eqref{eq:xi_inv} and
\eqref{eq:eta_inv}) to polar transits (i.e., the test particle is passing through
the negative or positive $z$ axis). But, according to the definition of parabolic coordinates, $\xi$ and $\eta$
are strictly non-negative quantities and thus the trajectories in the phase planes cannot
enter the regions $\xi < 0$ and $\eta < 0$. In order to solve this apparent contradiction
it can be shown how, in correspondence of a transit through $\xi = 0$ or $\eta = 0$, the corresponding
momentum ($p_\xi$ or $p_\eta$) switches discontinuously its sign  (and, concurrently, the azimuthal angle $\phi$
discontinuously changes by $\pm\pi$). In the phase plane, upon reaching the vertical axis from a positive
$\xi$ or $\eta$, the trajectory will be discontinuously reflected with respect to the horizontal axis,
and its evolution will proceed again towards positive $\xi$ or $\eta$.
Figure \ref{fig:bidimensional_case}
shows a selection of representative trajectories in the phase space for
$p_{\xi}\left(\xi\right)$ and $p_{\eta}\left(\eta\right)$ in the bidimensional case.


We proceed now to determine the explicit solutions for $\xi\left(\tau\right)$
and $\eta\left(\tau\right)$ in the three-dimensional case. We will focus on the study of the solution
for $\xi$, as the solution for $\eta$ differs only by notation. We will then
use $\xi\left(\tau\right)$ and $\eta\left(\tau\right)$ to determine the solution
for $\phi\left(\tau\right)$.

\section{Solution by quadrature}

The integration of eq. \eqref{eq:xi_diffeq} yields
\begin{equation}
\int_{0}^{\tau}du=\pm\int_{\xi_{0}}^{\xi}\frac{udu}{\sqrt{\varepsilon u^{6}+2hu^{4}+2\alpha_{1}u^{2}-p_{\phi}^{2}}},
\label{eq:tau_inverse}
\end{equation}
where the initial fictitious time has been set to zero\footnote{Note that one can always set the initial
fictitious time to zero, as the relation between real and fictitious time is differential -- see eq. \eqref{eq:fic_time}.
}, $u$ is a
dummy integration variable and $\xi_{0}$ is the initial value of
$\xi$. Before proceeding, we need to discuss briefly the nature of the
sign ambiguity in this formula.

The left-hand side of eq. \eqref{eq:tau_inverse}
represents the fictitious time needed by the dynamical system to evolve from the initial coordinate $\xi_0$ to an
arbitrary coordinate $\xi$. As pointed out in the previous section, all phase plots
are symmetric with respect to the horizontal axis, and thus,
along a trajectory in phase space, each coordinate $\xi$ will be visited twice:
once with a positive $p_\xi$ coordinate, and once with a negative $p_\xi$ coordinate\footnote{In the particular
case in which $p_\xi\left( \xi \right)$ has no real roots, there will be no sign ambiguity: $p_\xi$
will always be positive or negative, and the sign can be chosen once and for all in accordance
with the initial sign of $p_\xi$.}.
It follows that we can choose either sign in \eqref{eq:tau_inverse}, and the
left-hand side will then represent the evolution time along a portion of trajectory
in which $p_\xi$ remains positive ($+$) or negative ($-$).

Changing now integration variable in the right-hand side of eq. \eqref{eq:tau_inverse} to
\begin{equation}
s=\frac{1}{2}u^{2},
\end{equation}
we can rewrite the equation as
\begin{align}
\tau&=\pm\int_{\frac{1}{2}\xi_{0}^{2}}^{\frac{1}{2}\xi^{2}}\frac{ds}{\sqrt{8\varepsilon s^{3}+8hs^{2}+4\alpha_{1}s-p_{\phi}^{2}}} \label{eq:tau_t}\\
&=\pm\int_{\frac{1}{2}\xi_{0}^{2}}^{\frac{1}{2}\xi^{2}}\frac{ds}{\sqrt{f_{\xi}\left(s\right)}},
\end{align}
where $f_{\xi}\left(s\right)$ is a cubic polynomial in $s$. The
integral in this expression is an elliptic integral, which can be
computed and inverted to yield $\xi^{2}$ as function of $\tau$ using
a formula by Weierstrass \citep[see][\S 20.6]{whittaker_course_1927}.
After electing
\begin{align}
f_{\xi}\left(s\right) & =a_{4}+4a_{3}s+6a_{2}s^{2}+4a_{1}s^{3},
\end{align}
and defining
\begin{align}
g_{2} & =-4a_{1}a_{3}+3a_{2}^{2},\label{eq:g2_def}\\
g_{3} & =2a_{1}a_{2}a_{3}-a_{2}^{3}-a_{1}^{2}a_{4},\label{eq:g3_def}\\
\wp_{\xi}\left(\tau\right) & \equiv\wp\left(\tau;g_{2},g_{3}\right),
\end{align}
where $\wp\left(\tau;g_{2},g_{3}\right)$ is a Weierstrass elliptic
function defined in terms of the invariants $g_{2}$ and $g_{3}$ (see \citet{whittaker_course_1927}, Chapter XX, and \citet{abramowitz_handbook_1964}, Chapter 18),
the evolution of $\xi^{2}$ in fictitious time is given by
\begin{align}
\xi^{2} & =\xi_{0}^{2}+\frac{1}{\left[\wp_{\xi}\left(\tau\right)-\frac{1}{24}f_{\xi}^{\prime\prime}\left(\frac{\xi_{0}^{2}}{2}\right)\right]^{2}}\nonumber \\
 & \quad\cdot\left\{ \frac{1}{2}f_{\xi}^{\prime}\left(\frac{\xi_{0}^{2}}{2}\right)\left[\wp_{\xi}\left(\tau\right)-\frac{1}{24}f_{\xi}^{\prime\prime}\left(\frac{\xi_{0}^{2}}{2}\right)\right]\vphantom{\sqrt{f_{\xi}\left(\frac{\xi_{0}^{2}}{2}\right)}}\right.\nonumber \\
 & \quad\left.+\frac{1}{24}f_{\xi}\left(\frac{\xi_{0}^{2}}{2}\right)f_{\xi}^{\prime\prime\prime}\left(\frac{\xi_{0}^{2}}{2}\right)\pm\sqrt{f_{\xi}\left(\frac{\xi_{0}^{2}}{2}\right)}\wp_{\xi}^{\prime}\left(\tau\right)\right\}.\label{eq:xi_tau}
\end{align}
Here the $\pm$ sign represents the sign ambiguity discussed earlier,
and the derivatives of $f_\xi$ are calculated with respect to the polynomial variable,
while the derivative $\wp_\xi^\prime$ is calculated with respect to $\tau$. $\wp_\xi^\prime$ is related to $\wp$ via the relation
\begin{equation}
\left[\wp_\xi^{\prime}\left( z \right)\right] ^2 = 4 \wp_\xi^3\left( z \right) - g_2 \wp_\xi\left( z \right) - g_3 \label{eq:Pprime_rel}
\end{equation}
\citep[][eq. 18.1.6]{abramowitz_handbook_1964}. If $\xi_{0}^{2}/2$ is chosen as a root $\xi_{r}^{2}/2$
of $f_{\xi}$, then $f_\xi\left( \xi_r^2/2 \right) = 0$ and eq. \eqref{eq:xi_tau} simplifies to
\begin{equation}
\xi^{2}=\xi_{r}^{2}+\frac{1}{2}\frac{f_{\xi}^{\prime}\left(\frac{\xi_{r}^{2}}{2}\right)}{\wp_{\xi}\left(\tau-\tau_{\xi}\right)-\frac{1}{24}f_{\xi}^{\prime\prime}\left(\frac{\xi_{r}^{2}}{2}\right)},
\label{eq:xi_simpl}
\end{equation}
where $\tau_{\xi}$ is the fictitious time for which $\xi$ assumes
the value $\xi_{r}$. The analogous expressions for $\eta$ are
\begin{align}
\eta^{2} & =\eta_{0}^{2}+\frac{1}{\left[\wp_{\eta}\left(\tau\right)-\frac{1}{24}f_{\eta}^{\prime\prime}\left(\frac{\eta_{0}^{2}}{2}\right)\right]^{2}}\nonumber \\
 & \quad\cdot\left\{ \frac{1}{2}f_{\eta}^{\prime}\left(\frac{\eta_{0}^{2}}{2}\right)\left[\wp_{\eta}\left(\tau\right)-\frac{1}{24}f_{\eta}^{\prime\prime}\left(\frac{\eta_{0}^{2}}{2}\right)\right]\vphantom{\sqrt{f_{\eta}\left(\frac{\eta_{0}^{2}}{2}\right)}}\right.\nonumber \\
 & \quad\left.+\frac{1}{24}f_{\eta}\left(\frac{\eta_{0}^{2}}{2}\right)f_{\eta}^{\prime\prime\prime}\left(\frac{\eta_{0}^{2}}{2}\right)\pm\sqrt{f_{\eta}\left(\frac{\eta_{0}^{2}}{2}\right)}\wp_{\eta}^{\prime}\left(\tau\right)\right\} \label{eq:eta_tau}
\end{align}
and
\begin{equation}
\eta^{2}=\eta_{r}^{2}+\frac{1}{2}\frac{f_{\eta}^{\prime}\left(\frac{\eta_{r}^{2}}{2}\right)}{\wp_{\eta}\left(\tau-\tau_{\eta}\right)-\frac{1}{24}f_{\eta}^{\prime\prime}\left(\frac{\eta_{r}^{2}}{2}\right)}.
\label{eq:eta_simpl}
\end{equation}
The formul\ae{} \eqref{eq:xi_tau} and \eqref{eq:eta_tau} represent a general and complete closed-form solution for the squares $\xi^2$ and $\eta^2$ of the
parabolic coordinates $\xi$ and $\eta$. Since $\xi$ and $\eta$ are non-negative by definition, in order to recover the solution for
$\xi$ and $\eta$ it will be enough to take the principal square root of $\xi^2$ and $\eta^2$. The cartesian positions and velocities
can be reconstructed using eqs. \eqref{eq:x_para}-\eqref{eq:z_para}, where the derivatives of the parabolic coordinates
with respect to the real time can be computed by
inverting eqs. \eqref{eq:p_xi}-\eqref{eq:p_phi} (and by keeping in mind that $p_\xi$ and $p_\eta$ can be calculated by differentiating 
eqs. \eqref{eq:xi_tau} and \eqref{eq:eta_tau} with respect to $\tau$ -- see eqs. \eqref{eq:xi_diffeq} and \eqref{eq:eta_diffeq}).

For simplicity's sake, notational convenience and further analysis,
however, it is desirable to be able to use the simplified formul\ae{} \eqref{eq:xi_simpl} and \eqref{eq:eta_simpl} whenever possible.
To this end, we first note how, from the considerations presented in the previous section, the polynomials $f_\xi$ and $f_\eta$
will always have at least one positive real root, with the exception of the bidimensional case for $f_\xi$ with $\alpha_1 > 0$
(displayed in Figure \ref{fig:bidimensional_case} (b)). The roots $\xi_r$ and $\eta_r$ of the cubic polynomials $f_\xi$ and $f_\eta$
can be computed exactly using the general formul\ae{} for the roots of a cubic function.
Secondly, in order to determine $\tau_\xi$ (and, analogously, $\tau_\eta$)
we can use eq. \eqref{eq:tau_t} to write:
\begin{equation}
\tau_\xi=\pm\int_{\frac{1}{2}\xi_{0}^{2}}^{\frac{1}{2}\xi_r^{2}}\frac{ds}{\sqrt{8\varepsilon s^{3}+8hs^{2}+4\alpha_{1}s-p_{\phi}^{2}}}.
\label{eq:depressed_transform}
\end{equation}
Following then \citet[][eqs. (A7)--(A13)]{byrd_handbook_1971}, we introduce the Tschirnhaus transformation \citep{cayley_tschirnhausens_1861}
\begin{equation}
s = \sqrt[3]{\frac{1}{2\varepsilon}}s_1 - \frac{1}{3}\frac{h}{\varepsilon}
\end{equation}
in order to reduce the polynomial $f_\xi$ to a depressed cubic:
\begin{equation}
\tau_\xi=\pm\sqrt[3]{\frac{1}{2\varepsilon}}\int_{\sqrt[3]{2\varepsilon}\left(\frac{1}{2}\xi_0^2+\frac{1}{3}\frac{h}{\varepsilon}\right)}^
{\sqrt[3]{2\varepsilon}\left(\frac{1}{2}\xi_r^2+\frac{1}{3}\frac{h}{\varepsilon}\right)}
\frac{ds_1}{
\sqrt{4s_{1}^{3}-h_2s_{1}-h_3}},
\end{equation}
where
\begin{align}
h_2 & = \sqrt[3]{\frac{1}{2\varepsilon}}\left(\frac{8}{3}\frac{h^{2}}{\varepsilon}-4\alpha_{1}\right), \\
h_3 & = \frac{4}{3}\frac{\alpha_{1}h}{\varepsilon}-\frac{16}{27}\frac{h^{3}}{\varepsilon^{2}}+p_{\phi}^{2}.
\end{align}
The integral can now be split into two separate Weierstrass normal elliptic integrals of the first kind,
\begin{multline}
\tau_\xi=\pm\sqrt[3]{\frac{1}{2\varepsilon}}\left[
\int_{\sqrt[3]{2\varepsilon}\left(\frac{1}{2}\xi_0^2+\frac{1}{3}\frac{h}{\varepsilon}\right)}^{\infty}
\frac{ds_1}{\sqrt{4s_{1}^{3}-h_2s_{1}-h_3}}
\right.\\
\left.
-\int_{\sqrt[3]{2\varepsilon}\left(\frac{1}{2}\xi_r^2+\frac{1}{3}\frac{h}{\varepsilon}\right)}^{\infty}
\frac{ds_1}{\sqrt{4s_{1}^{3}-h_2s_{1}-h_3}}
\right],
\end{multline}
and solved in terms of the inverse Weierstrass elliptic function $\wp^{-1}$ as
\begin{multline}
\tau_\xi=\pm\sqrt[3]{\frac{1}{2\varepsilon}}\left\{
\wp^{-1}\left[\sqrt[3]{2\varepsilon}\left(\frac{1}{2}\xi_0^2+\frac{1}{3}\frac{h}{\varepsilon}\right);h_2,h_3\right]\right.\\
\left.-\wp^{-1}\left[\sqrt[3]{2\varepsilon}\left(\frac{1}{2}\xi_r^2+\frac{1}{3}\frac{h}{\varepsilon}\right);h_2,h_3\right]
\right\}.\label{eq:tau_xi}
\end{multline}
The corresponding formula for $\tau_\eta$ can be obtained by switching $\varepsilon$ to $-\varepsilon$ and $\alpha_1$ to $\alpha_2$.
It must be noted though that in this formula there are ambiguities regarding the computation of the inverse
Weierstrass elliptic function, as $\wp^{-1}\left(z\right)$ is a multivalued function\footnote{Not only $\wp\left(z\right)$
is doubly periodic in $z$, but even within its fundamental periods it assumes all complex values twice \citep{whittaker_course_1927}.}.
The values of $\wp^{-1}$ in eq. \eqref{eq:tau_xi} have then to be chosen appropriately in order to yield the correct result \citep[as explained, e.g., in][]{hoggatt_inverse_1955}.
As an alternative, it is possible to compute directly the integral in eq. \eqref{eq:depressed_transform} in terms of Legendre elliptic
integrals using known formul\ae{} \citep[e.g.,][\S 3.131 and \S 3.138]{gradshtein_table_2007}.

The solution for the third coordinate $\phi$ can now be computed directly by integrating eq. \eqref{eq:phi_diffeq} with respect to $\tau$:
\begin{equation}
\int_{\phi_0}^\phi du = p_\phi\left[\int_0^\tau\frac{du}{\xi^2\left( u \right)} + \int_0^\tau\frac{du}{\eta^2\left( u \right)}\right].
\label{eq:phi_inteq}
\end{equation}
It is easier to tackle the calculation via the simplified formul\ae{} \eqref{eq:xi_simpl} and \eqref{eq:eta_simpl}. The integrals
on the right-hand side of eq. \eqref{eq:phi_inteq} are then in a form which can be solved through a formula involving $\wp^\prime$, $\wp^{-1}$
and the Weierstrass $\sigma$ and $\zeta$ functions (see \citet{jules_tannery_elements_1893},
chapter CXII, and \citet{gradshtein_table_2007}, \S 5.141.5):
\begin{equation}
\int \frac{\wp\left( u \right) + \beta}{\gamma\wp\left( u \right) + \delta}du =
\frac{u}{\gamma}+\frac{\delta-\beta\gamma}{\gamma^2\wp^\prime\left(v\right)}\left[\ln\frac{\sigma
\left( u+v \right)}{\sigma\left( u-v \right)} - 2u\zeta\left( v \right)\right],
\label{eq:phi_integral}
\end{equation}
where $v=\wp^{-1}\left(-\delta / \gamma\right)$.
In this case, the multivalued character of $\wp^{-1}$ does not matter: it can be verified via the reduction formul\ae{}
of the Weierstrassian functions \citep[][\S 18.2]{abramowitz_handbook_1964} that \emph{any} value of $v$ such
that $\wp\left( v \right) = -\delta/\gamma$ will yield the same result in the right-hand side of eq. \eqref{eq:phi_integral}.
The final result for $\phi$ is then\footnote{There is an insidious technical difficulty in the direct use of formula \eqref{eq:phi_tau},
related to the multivalued character of the complex logarithm. The issue is presented and addressed in Appendix \ref{sec:complex_log}.}:
\begin{multline}
\phi = \phi_0 + 2p_\phi\left\{\tau\left(\frac{1}{\gamma_\xi}+\frac{1}{\gamma_\eta}\right)+\frac{\delta_\xi-\beta_\xi\gamma_\xi}{\gamma_\xi^2\wp_\xi^\prime\left(u_\xi\right)}\right.\\
\cdot\left[\ln\frac{\sigma_\xi\left(\tau-\tau_\xi+u_\xi\right)}{\sigma_\xi\left(\tau-\tau_\xi-u_\xi\right)}
-\ln\frac{\sigma_\xi\left(-\tau_\xi+u_\xi\right)}{\sigma_\xi\left(-\tau_\xi-u_\xi\right)}-2\tau\zeta_\xi\left( u_\xi \right)\right]\\
+\frac{\delta_\eta-\beta_\eta\gamma_\eta}{\gamma_\eta^2\wp_\eta^\prime\left(u_\eta\right)}\\
\left.\cdot\left[\ln\frac{\sigma_\eta\left(\tau-\tau_\eta+u_\eta\right)}{\sigma_\eta\left(\tau-\tau_\eta-u_\eta\right)}
-\ln\frac{\sigma_\eta\left(-\tau_\eta+u_\eta\right)}{\sigma_\eta\left(-\tau_\eta-u_\eta\right)}
-2\tau\zeta_\eta\left( u_\eta \right)\right]
\vphantom{\frac{\tau}{\gamma_\xi}+\frac{\delta_\xi-\beta_\xi\gamma_\xi}{\gamma_\xi^2\wp_\xi^\prime\left(u_\xi\right)}}
\right\},\label{eq:phi_tau}
\end{multline}
where the following constants have been defined for notational convenience:
\begin{align}
\beta_\xi & = -\frac{1}{24}f_{\xi}^{\prime\prime}\left(\frac{\xi_{r}^{2}}{2}\right),&\beta_\eta & = -\frac{1}{24}f_{\eta}^{\prime\prime}\left(\frac{\eta_{r}^{2}}{2}\right),\\
\gamma_\xi & = 2\xi_r^2, & \gamma_\eta & = 2\eta_r^2,\\
\delta_\xi & = f_\xi^\prime\left(\frac{\xi_r^2}{2}\right)+2\xi_r^2\beta_\xi,&\delta_\eta & = f_\eta^\prime\left(\frac{\eta_r^2}{2}\right)+2\eta_r^2\beta_\eta,\\
u_\xi&=\wp_\xi^{-1}\left(-\frac{\delta_\xi}{\gamma_\xi}\right),&u_\eta&=\wp_\eta^{-1}\left(-\frac{\delta_\eta}{\gamma_\eta}\right).
\end{align}
Starting from eq. \eqref{eq:phi_tau}, we adopt the subscript notation $\sigma_\xi$ and $\zeta_\xi$ to indicate Weierstrass $\sigma$ and $\zeta$ functions
defined in terms of the same invariants as $\wp_\xi$.
\section{The time equation}
\label{sec:time_equation}
The final step in the solution of the Stark problem is to establish an explicit connection between real and fictitious time.
To this end, we need to integrate eq. \eqref{eq:fic_time}:
\begin{equation}
dt=\left[\xi^{2}\left(\tau\right)+\eta^{2}\left(\tau\right)\right]d\tau.
\end{equation}
In the general case, according to eqs. \eqref{eq:xi_tau} and \eqref{eq:eta_tau}, the
exact solutions for $\xi^{2}\left(\tau\right)$ and $\eta^{2}\left(\tau\right)$ are of the form
\begin{equation}
A+B\wp^{\prime}\left(\tau\right),\label{eq:wp_rational}
\end{equation}
where A and B are rational functions of $\wp\left(\tau\right)$.
Then, according to the theory of elliptic functions, the antiderivative of \eqref{eq:wp_rational} can be calculated in terms of $\wp$, $\wp^\prime$, $\wp^{-1}$
and the Weierstrass $\sigma$ and $\zeta$ functions. The integration method, due to \citet[][chapter VII]{halphen_traite_1886} (and explained in detail
in \citet[][chapter VII]{greenhill_applications_1959}), involves the decomposition of $A$ and $B$ into separate fractions, resulting in the split of the integral
into fundamental forms that can be integrated using the Weierstrassian functions.

It is again easier to use the simplified solutions \eqref{eq:xi_simpl} and \eqref{eq:eta_simpl}, and thus obtain the time equation
\begin{align}
t &= \int_0^\tau\left[\xi^{2}\left(u\right)+\eta^{2}\left(u\right)\right]du \\
& = \left(\xi_r^2+\eta_r^2\right)\tau + \frac{1}{2}\int_0^\tau
\frac{f_{\xi}^{\prime}\left(\frac{\xi_{r}^{2}}{2}\right)}{\wp_{\xi}\left(u-\tau_{\xi}\right)
-\frac{1}{24}f_{\xi}^{\prime\prime}\left(\frac{\xi_{r}^{2}}{2}\right)}du\notag\\
&\quad +\frac{1}{2}\int_0^\tau
\frac{f_{\eta}^{\prime}\left(\frac{\eta_{r}^{2}}{2}\right)}{\wp_{\eta}\left(u-\tau_{\eta}\right)
-\frac{1}{24}f_{\eta}^{\prime\prime}\left(\frac{\eta_{r}^{2}}{2}\right)}du.\label{eq:tau_01}
\end{align}
The integrals appearing in eq. \eqref{eq:tau_01} are known and they can be computed directly. To this end, it would be tempting to
apply the formul\ae{} in \citet[][\S 5.141]{gradshtein_table_2007}. However, as it can be verified by direct substitution
using the exact solution of the cubic equation\footnote{Such a check is best performed using
a computer algebra tool. In this specific case, we used the Python library SymPy \citep{sympy}.}, $\frac{1}{24}f_{\xi}^{\prime\prime}\left(\frac{\xi_{r}^{2}}{2}\right)$ and
$\frac{1}{24}f_{\eta}^{\prime\prime}\left(\frac{\eta_{r}^{2}}{2}\right)$ are always roots of the characteristic cubic equations
\begin{equation}
4t^3 - g_2t - g_3 = 0,\label{eq:char_cubic}
\end{equation}
associated to $\wp_\xi$ and $\wp_\eta$. Consequently, the formul\ae{} in \citet[][\S 5.141]{gradshtein_table_2007} will be singular,
and we have to use instead the results in \citet[][\S CXII]{jules_tannery_elements_1893}, which yield the formula
\begin{equation}
\int\frac{du}{\wp\left(u\right)-e_i} = \frac{1}{g_2/4-3e_i^2}\left[ue_i + \zeta\left(u-\omega_i\right)\right].
\end{equation}
In this formula, the $e_i$ represents the three roots of the characteristic cubic equation, while the $\omega_i$ are defined
by the relation $e_i=\wp\left(\omega_i\right)$ (so that, following \citet[][eq. 18.3.1]{abramowitz_handbook_1964}, two of the $\omega_i$
are the fundamental half-periods of $\wp$ and the third one is the sum of the fundamental half-periods). The solution of eq. \eqref{eq:tau_01}
is thus:
\begin{multline}
t = \left(\xi_r^2+\eta_r^2\right)\tau+\frac{1}{2}\frac{f_{\xi}^{\prime}\left(\frac{\xi_{r}^{2}}{2}\right)}{g_{2,\xi}/4-3e_{i,\xi}^2}\cdot
\\\left[\tau e_{i,\xi} + \zeta_\xi\left(\tau-\tau_\xi-\omega_{i,\xi}\right)-\zeta_\xi\left(-\tau_\xi-\omega_{i,\xi}\right)\right]+\\
\frac{1}{2}\frac{f_{\eta}^{\prime}\left(\frac{\eta_{r}^{2}}{2}\right)}{g_{2,\eta}/4-3e_{i,\eta}^2}\cdot\\
\left[\tau e_{i,\eta} + \zeta_\eta\left(\tau-\tau_\eta-\omega_{i,\eta}\right)-\zeta_\eta\left(-\tau_\eta-\omega_{i,\eta}\right)\right].
\label{eq:time_equation}
\end{multline}
This equation can be considered as the equivalent of Kepler's equation for the Stark problem. Similarly to the two-body problem, it is constituted
of a linear part modulated by two quasi-periodic transcendental parts (with the Weierstrass $\zeta$ function
replacing the sine function appearing in Kepler's equation). In this sense, the fictitious time $\tau$ can be
regarded as a kind of eccentric anomaly for the Stark problem. According to eq. \eqref{eq:fic_time}, the time equation is a monotonic function
and its inversion can thus be achieved numerically using standard techniques (Newton-Raphson, bisection, etc.).
\section{Analysis of the results}
After having determined the full formal solution of the Stark problem in the previous sections, we now turn our attention
to the interpretation of the results.

Before proceeding, we first need to point out how our solution to the Stark problem, as developed in the previous sections, is directly applicable
to the three-dimensional case, but not in general to all bidimensional cases.
As explained in \S \ref{subsec:poly_study}, in certain bidimensional cases (specifically, when
the constant of motion $\alpha_1$ is positive) the polynomial $p_\xi\left(\xi\right)$
might have no real zeroes, and thus the simplified formula \eqref{eq:xi_simpl} cannot be used.
While in this case it is still possible to proceed to a complete solution via the full formula \eqref{eq:xi_tau}
in conjunction with the general theory for the integration of rational functions of elliptic functions
(see \citet[][chapter VII]{halphen_traite_1886} and \citet[][chapter VII]{greenhill_applications_1959}),
the resulting expressions for $\phi\left(\tau\right)$ and $t\left(\tau\right)$
will be more complicated than the formul\ae{} obtained for the three-dimensional case.

An additional complication in the bidimensional case is the presence of the discontinuity discussed in
\S \ref{subsec:poly_study}. In correspondence of a polar transit, either $p_\xi$ or $p_\eta$ will switch
sign. This discontinuity must be taken into account in the computation and inversion of the integral
\eqref{eq:tau_t}, and ultimately it has the effect of introducing a branching in the solutions for
$\xi\left(\tau\right)$ and $\eta\left(\tau\right)$.

\subsection{Quasi-periodicity and periodicity}
Our solution to the Stark problem is based on the Weierstrass elliptic and related functions.
Without giving a full account of the theory of the Weierstrassian functions (for which we refer to standard textbooks such as
\citet{whittaker_course_1927}), we will recall here briefly a few fundamental notions\footnote{It is interesting
to note that the study of the Weierstrassian formalism for the theory of elliptic functions is today no longer part of the typical background of physicists and engineers.
Recently, the Weierstrassian formalism has been successfully applied to dynamical studies in General Relativity
\citep[e.g.,][]{hackmann_analytical_2010,scharf_schwarzschild_2011,gibbons_application_2012,biscani_first-order_2013}.}. To this end, we will employ the notation
of \citet[][chapter 18]{abramowitz_handbook_1964}.

The elliptic function $\wp\left( z;g_2,g_3\right)$ is a doubly-periodic complex-valued function of a complex variable $z$ defined in terms of two
complex parameters $g_2$ and $g_3$, called \emph{invariants}. The complex primitive half-periods $\omega$ and $\omega^\prime$
of $\wp$ can be related to the invariants via formul\ae{}
involving elliptic integrals and the roots $e_1$, $e_2$ and $e_3$ of the characteristic cubic equation
\begin{equation}
4t^3-g_2t-g_3 = 0 \label{eq:cubic_wp}
\end{equation}
\citep[e.g., see][\S 18.9]{abramowitz_handbook_1964}. The sign of the \emph{modular discriminant}
\begin{equation}
\Delta = g_2^3-27g_3^2
\end{equation}
determines the nature of the roots $e_1$, $e_2$ and $e_3$. In the case of the Stark problem, the invariants are by definition real
(see eqs. \eqref{eq:g2_def}--\eqref{eq:g3_def}),
and thus the $(\omega,\omega^\prime)$ pairs can be chosen as (real, imaginary) or complex conjugate
(depending on the sign of $\Delta$).
It is known from the theory of elliptic
functions that there actually exist infinite pairs of fundamental half-periods
for $\wp$, related to each other via integral linear combinations
with unitary determinant \citep[][\S 79]{hancock_lectures_1910}. We can then always
introduce two new half-periods $\omega_R$ (the \emph{real} period) and
$\omega_C$ (the \emph{complex} period) such that $\omega_R$ is real and positive,
and $\omega_C$ complex with positive imaginary part. The relation with the fundamental
half-periods $\omega$ and $\omega^\prime$ from \citet{abramowitz_handbook_1964} is
\begin{align}
\omega_R & = \omega + \delta \omega^\prime,\\
\omega_C & = \omega^\prime,
\end{align}
where $\delta = 0$ if $\Delta > 0$ and $\delta = 1$ if $\Delta < 0$.
Since we are interested in the behaviour of $\wp$ on the real axis (as $\tau$ is
a real-valued variable), we can then regard $\wp\left( \tau;g_2,g_3\right)$ as a singly-periodic
real-valued function of period $2\omega_R$.

It follows then straightforwardly from eqs. \eqref{eq:xi_tau}--\eqref{eq:eta_simpl} that $\xi\left( \tau \right)$ and $\eta\left( \tau \right)$
are both periodic in $\tau$ with periods that, in general, will be different. Conversely, from eq.
\eqref{eq:phi_tau}, it follows immediately that $\phi\left( \tau \right)$ is not periodic.
Indeed, $\phi\left( \tau \right)$ is a function of the form
\begin{equation}
f\left(\tau\right)=A+B\tau+C_\xi\ln\frac{\sigma_\xi\left(\tau + a_\xi\right)}{\sigma_\xi\left(\tau + b_\xi\right)}
+C_\eta\ln\frac{\sigma_\eta\left(\tau + a_\eta\right)}{\sigma_\eta\left(\tau + b_\eta\right)},\label{eq:generic_f_sigma}
\end{equation}
where $A$, $B$, $C$, $a$ and $b$ are constants.
It is now interesting to note that, according to eqs. \eqref{eq:fic_time} and \eqref{eq:phi_diffeq}, if $\xi$ and $\eta$ have
real half-periods $\omega_{R,\xi}$ and $\omega_{R,\eta}$ such that
\begin{equation}
\frac{\omega_{R,\xi}}{\omega_{R,\eta}}=\frac{n}{m},
\end{equation}
with $n$ and $m$ coprime natural numbers (or, in other words, $\omega_{R,\xi}$ and $\omega_{R,\eta}$ are commensurable),
then $d\phi/d\tau$ becomes a periodic function with period $T = 2m\omega_{R,\xi} = 2n\omega_{R,\eta}$.
Recalling the quasi-periodicity of $\sigma$ via the relation
\citep[][eq. 18.2.20]{abramowitz_handbook_1964}
\begin{multline}
\sigma\left(z+2M\omega+2N\omega^\prime\right) = \left(-1\right)^{M+N+MN}\sigma\left(z\right)\cdot \\
\mathrm{e}^{\left(z+M\omega+N\omega^\prime\right)\left[2M\zeta\left(\omega\right)+2N\zeta\left(\omega^\prime\right)\right]},
\end{multline}
with $M,N\in\mathbb{Z}$, we can then write for eq. \eqref{eq:generic_f_sigma}
\begin{multline}
f\left(\tau + T\right) = A+B\tau+C_\xi\ln\frac{\sigma_\xi\left(\tau + a_\xi\right)}{\sigma_\xi\left(\tau + b_\xi\right)}
+C_\eta\ln\frac{\sigma_\eta\left(\tau + a_\eta\right)}{\sigma_\eta\left(\tau + b_\eta\right)}\\
+ BT + 2 m C_\xi\left(a_\xi - b_\xi \right)\zeta_\xi\left(\omega_{R,\xi}\right)\\
+ 2 n C_\eta\left(a_\eta - b_\eta \right)\zeta_\eta\left(\omega_{R,\eta}\right),
\end{multline}
or, more succinctly,
\begin{equation}
f\left(\tau + T\right) = f\left(\tau\right) + D,
\end{equation}
where $D$ is the constant
\begin{multline}
D = BT + 2 m C_\xi\left(a_\xi - b_\xi \right)\zeta_\xi\left(\omega_{R,\xi}\right)\\
+ 2 n C_\eta\left(a_\eta - b_\eta \right)\zeta_\eta\left(\omega_{R,\eta}\right).
\end{multline}
Thus, if $\xi\left( \tau \right)$ and $\eta\left( \tau \right)$
have commensurable periods, $\phi\left( \tau \right)$ is an arithmetic quasi-periodic function of
$\tau$. The geometric meaning of this quasi-periodicity is that, after a quasi-period $T$,
the test particle will be in a position that results from a rotation around the $z$ axis of
the original position. The particle's trajectory will thus draw a rotationally-symmetric figure in space.

\begin{figure*}
\includegraphics[width=1\textwidth]{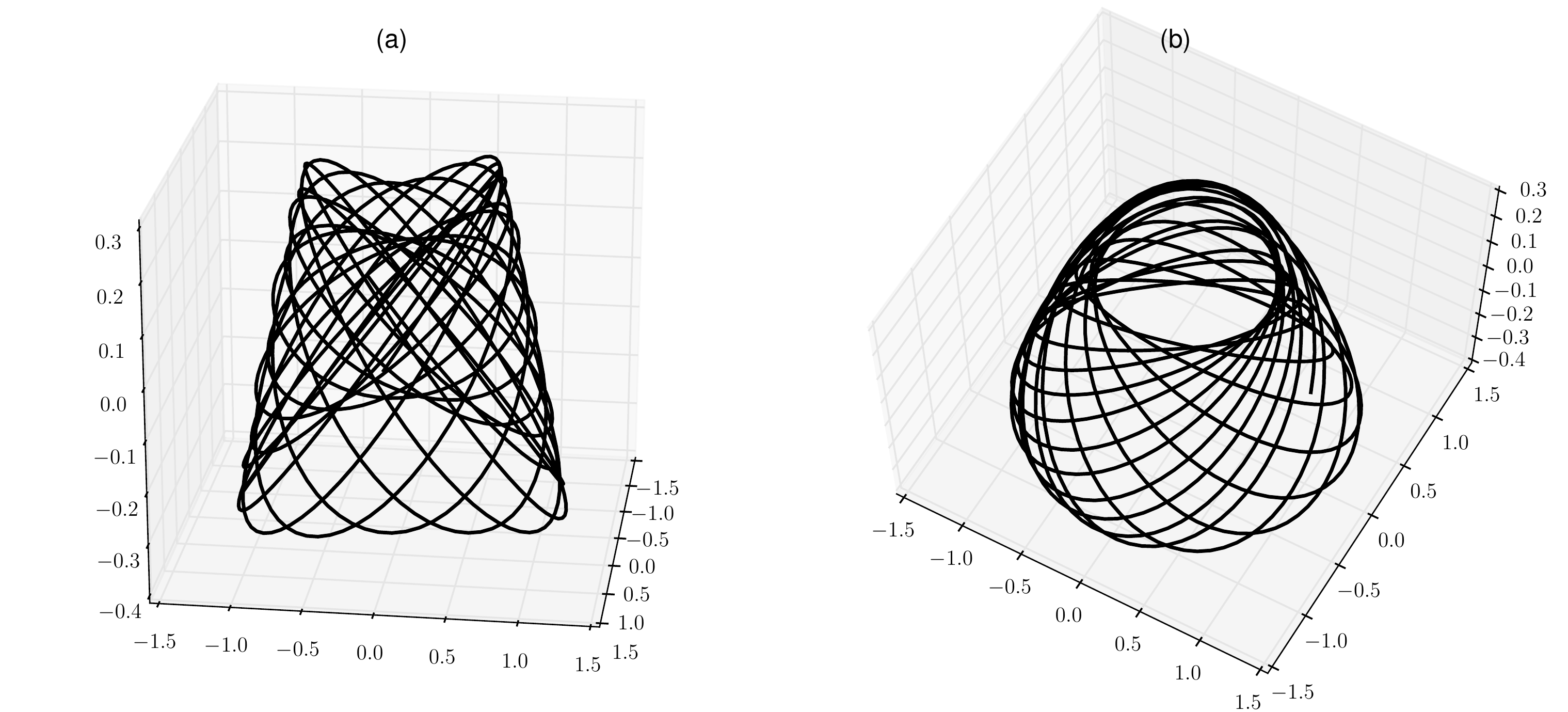}

\caption{Three-dimensional plots of a representative quasi-periodic orbit, seen from the side (a) and from the top (b). In this specific case,
the periods of $\xi$ and $\eta$ in fictitious time are in a ratio of $6/5$ within an accuracy of $\sim 10^{-11}$.
\label{fig:quasi_periodic}}
\end{figure*}

Quasi-periodic orbits can be found via a numerical search for a set of initial conditions
and constant acceleration field $\varepsilon$ that satisfies the commensurability relation on the periods of $\xi$ and $\eta$.
The numerical search can be setup as the minimisation of the function $\left(m\omega_{R,\xi}-n\omega_{R,\eta}\right)^2$ for
two chosen coprime integers $n$ and $m$. A representative quasi-periodic orbit found this way using the PaGMO optimiser
\citep{biscani_global_2010} is displayed in Figure \ref{fig:quasi_periodic}.

\begin{figure*}
\includegraphics[width=1\textwidth]{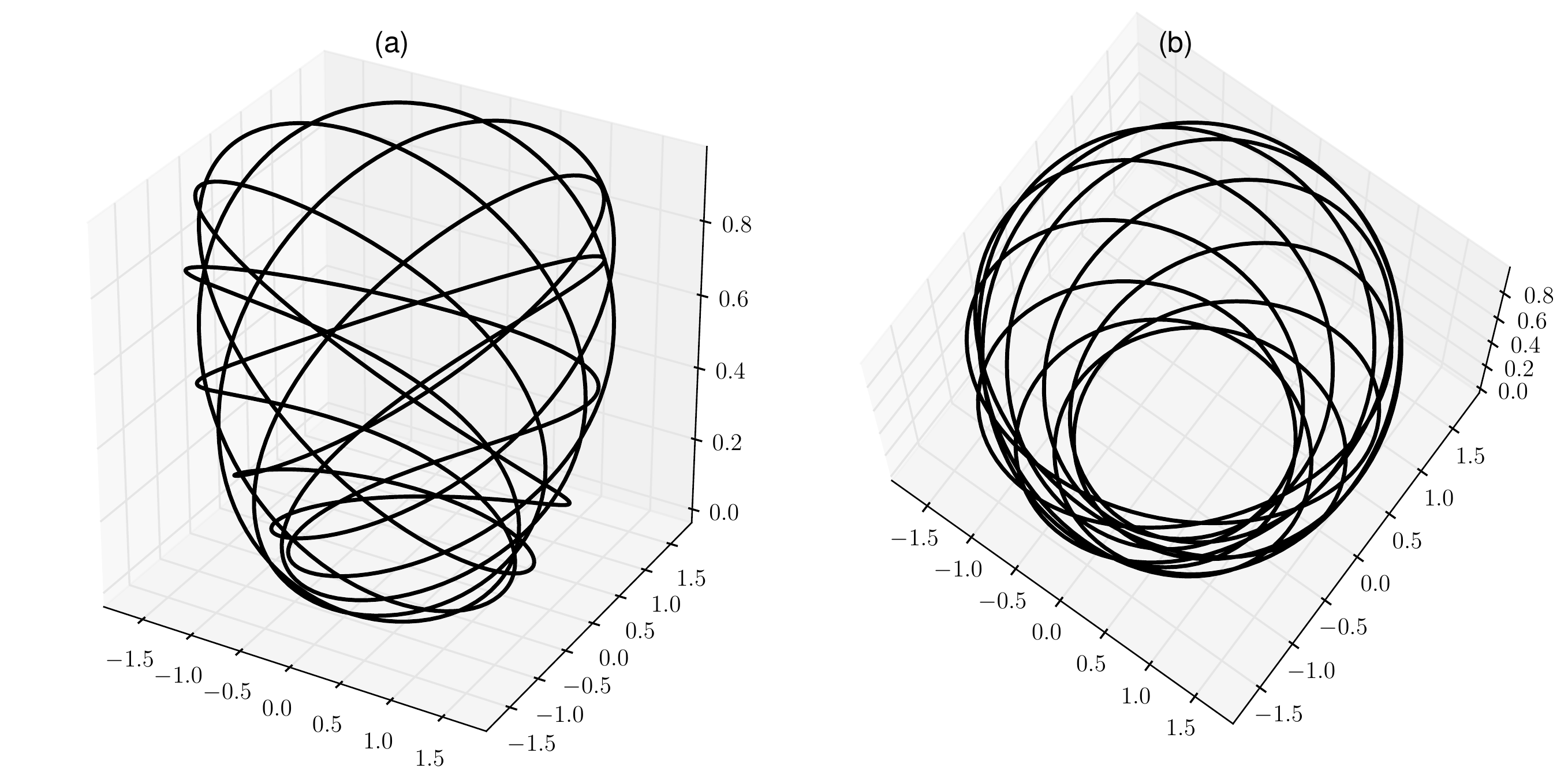}

\caption{Three-dimensional plots of a representative periodic orbit ($n=1$, $m = 2$, $p = 7$), seen from the side (a) and from the top (b). One period of the trajectory is displayed. In this specific
case, the trajectory is closed at the end of one period with an accuracy of $\sim 10^{-5}$.
\label{fig:periodic}}
\end{figure*}

Periodic orbits can also be found in a similar way by imposing the additional condition $p \phi(T) = 2\pi$,
where $p \in \mathbb Z$. For any triplet of $\left( n,m,p \right)$ integers, one has then to solve numerically an optimisation problem that
yields periodic orbits such as the one displayed in Figure \ref{fig:periodic} for a case $n=1$, $m = 2$, and $p = 7$.

\subsection{Bound and unbound orbits}
\begin{figure*}
\includegraphics[width=1\textwidth]{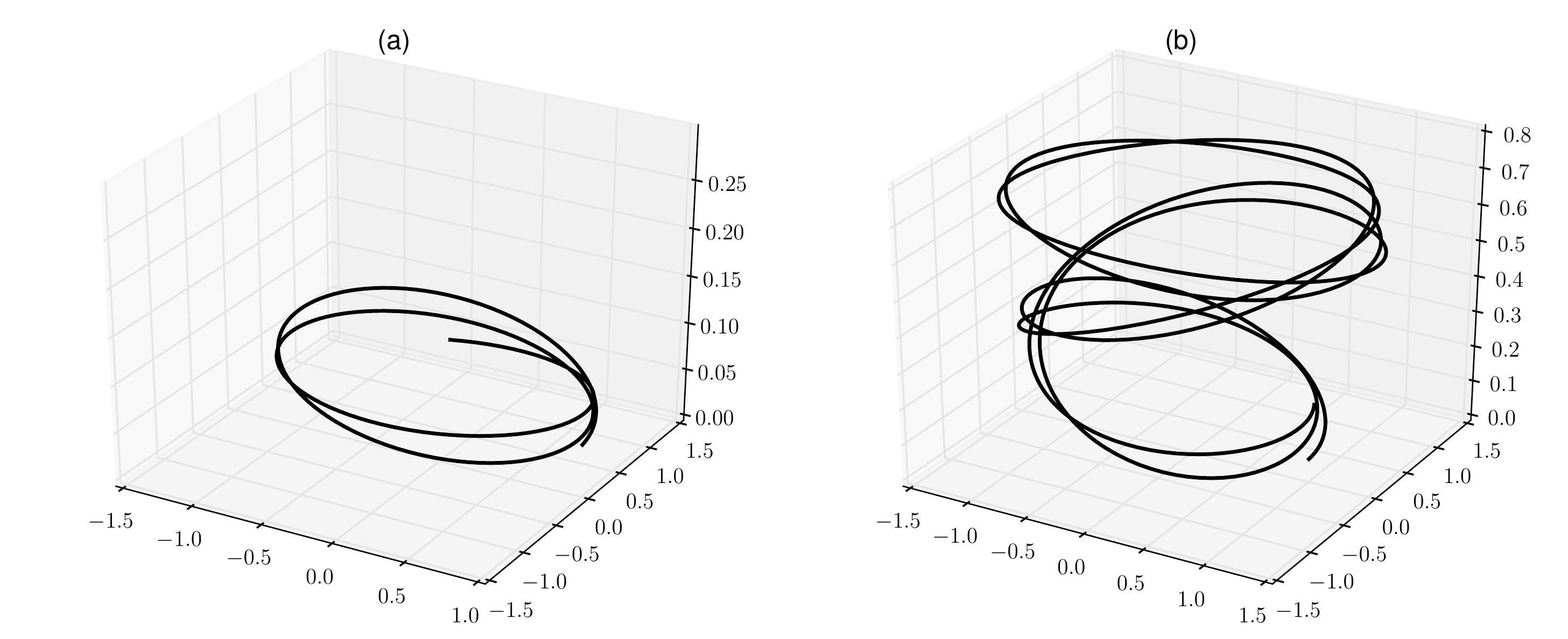}

\caption{Three-dimensional plots of two representative bound orbits sharing the same initial conditions but with different values for the constant acceleration field.
The initial condition corresponds (in absence of the external acceleration field) to a quasi-circular Keplerian orbit lying close to the $xy$ plane.
The acceleration field is weaker in (a), whereas in (b) it is close to the critical value for which the orbit becomes unbound.
\label{fig:plot_3d}}
\end{figure*}

The solution of the Stark problem in terms of the Weierstrassian functions allows to determine the conditions under which
the motion is bound. As we have seen in the previous sections, the parabolic coordinate $\eta$ is always bound, whereas $\xi$
can be either bound or unbound. From the general solution \eqref{eq:xi_tau}, it is easily deduced that the formula for $\xi\left(\tau\right)$
has a pole (and thus $\xi$ is unbound) when the denominator is zero, i.e., under the condition
\begin{equation}
\wp_{\xi}\left(\tau\right)-\frac{1}{24}f_{\xi}^{\prime\prime}\left(\frac{\xi_{0}^{2}}{2}\right) = 0.
\end{equation}
Recalling now that $\wp_\xi\left(\tau\right)$ is analytical everywhere except at the poles (where it behaves like $1/\tau^2$ around $\tau = 0$),
it can be deduced from the properties of parity and periodicity that $\wp_\xi\left(\tau\right)$ must have a global minimum within the real
period $2\omega_R$. Moreover, since $\wp_\xi$ satisfies the differential equation \eqref{eq:Pprime_rel},
the condition for the existence of a stationary point is
\begin{equation}
\wp_\xi\left(\tau\right) = e_i,
\end{equation}
where $e_i$ represents the roots of the cubic equation \eqref{eq:cubic_wp}.
It is known \citep[][eq. 18.3.1]{abramowitz_handbook_1964} that $\wp_\xi\left(\omega_i\right) = e_i$, where
\begin{align}
\omega_1 &= \omega,\\
\omega_2 &= \omega + \omega^\prime,\\
\omega_3 &= \omega^\prime,
\end{align}
which implies that the global minimum of $\wp_\xi\left(\tau\right)$ is in correspondence of $\tau = \omega_R$.
We can then conclude that the condition for bound motion is
\begin{equation}
e_R > \frac{1}{24}f_{\xi}^{\prime\prime}\left(\frac{\xi_{0}^{2}}{2}\right),
\end{equation}
where we have denoted with $e_R$ the root of the cubic equation \eqref{eq:cubic_wp} for which $\wp_\xi\left(\omega_R\right) = e_R$.
Figure \ref{fig:plot_3d} displays the evolution of two representative bound orbits in the three-dimensional space.

\begin{figure*}
\includegraphics[width=1\textwidth]{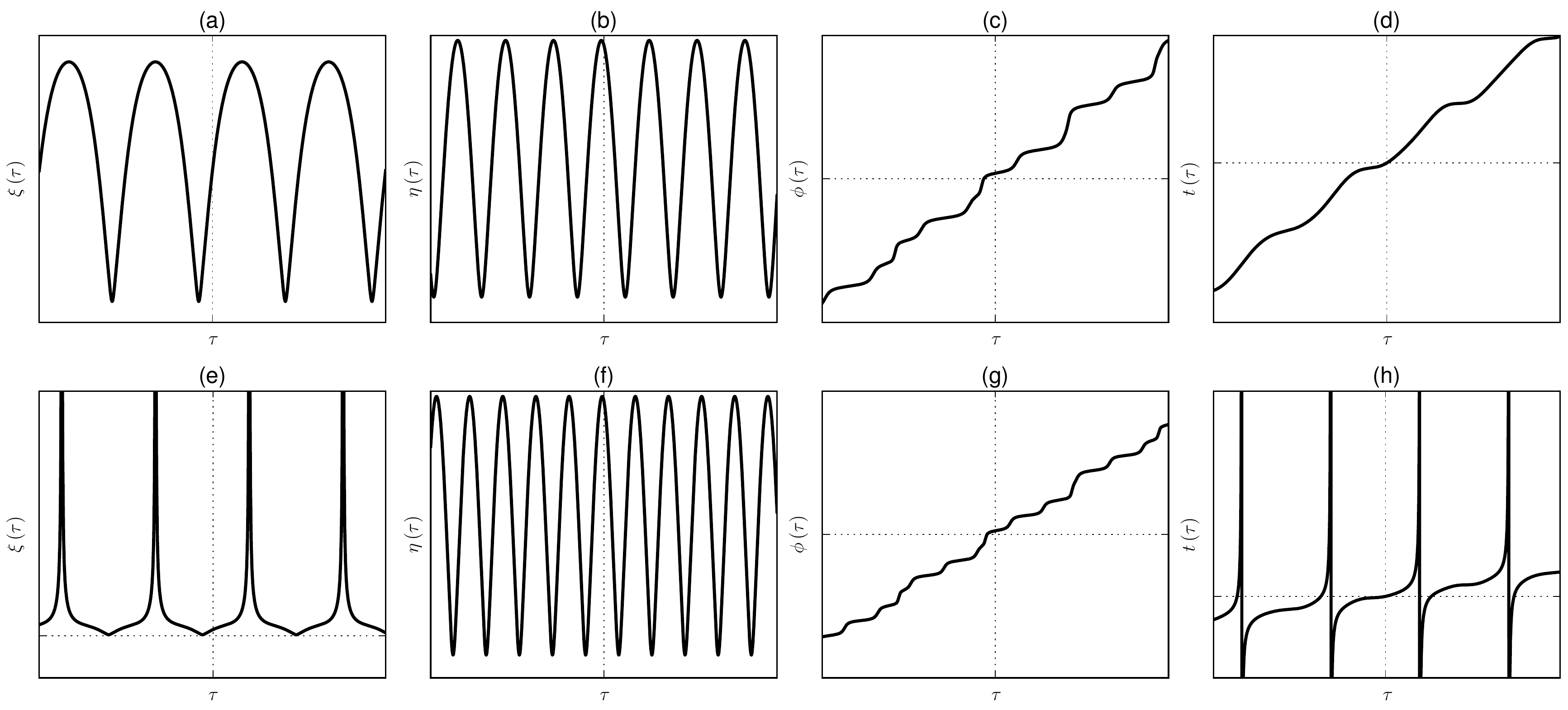}

\caption{Representative plots of the evolution in fictitious time $\tau$ of the parabolic coordinates $\xi$, $\eta$
and $\phi$ and of the real time $t$ in a bound ((a)-(d), first row) and an unbound ((e)-(h), second row) orbit. In the unbound case, the $\xi$ coordinate (e) and the real
time $t$ (h) reach infinity in a finite amount of fictitious time.\label{fig:bound_vs_unbound}}
\end{figure*}

Figure \ref{fig:bound_vs_unbound} displays the evolution in $\tau$ of the parabolic coordinates and of the real time $t$
in a bound and an unbound case. It is interesting to note that in the unbound case only $\xi$ and $t$ present vertical asymptotes, whereas
$\eta$ and $\phi$ assume finite values when $\xi$ and $t$ go to infinity. With respect to the evolution in real time $t$, this means
that $\eta$ and $\phi$ tend asymptotically to finite values for $t\to \infty$. At infinity, the trajectory of the test particle is determined solely by the
constant acceleration field and will thus be a parabola. The plane in which such asymptotic parabola lies is perpendicular to the $xy$ plane
and its orientation is determined by the value to which the azimuthal angle $\phi$ tends asymptotically (which can be determined exactly
by calculating the value of $\phi$ at the end of one period in fictitious time). This result could prove to be particularly useful in the design of
powered planetary kicks (or flybys), a technique vastly used in modern interplanetary trajectory design \citep{danby_fundamentals_1988}.
Planetary kicks are traditionally designed
assuming an unperturbed hyperbolic motion around a certain planet. The outgoing conditions are then simply determined by the analytical expression
governing Keplerian motion (i.e., a rotation of the hyperbolic access velocity). A different type of powered flyby can be considered, in which
the spacecraft thrusts continuously in a fixed inertial direction. In such a case, and ignoring the fuel mass loss,
the spacecraft conditions at infinity (i.e., when leaving the planet's sphere of influence) can be determined exactly by
a fully-analytical solution such as the one presented here.

\subsection{Equilibrium points and displaced circular orbits}
We turn now our attention to the analysis of the equilibrium points of the Stark problem. It is useful to consider initially the Hamiltonian in
cartesian coordinates and real time $t$ resulting from the Lagrangian \eqref{eq:orig_lagr}. The equations of motion are, trivially,
\begin{align}
\frac{dx}{dt}&=v_x,&\frac{dv_x}{dt}&=-\frac{\mu x}{r^3},\\
\frac{dy}{dt}&=v_y,&\frac{dv_y}{dt}&=-\frac{\mu y}{r^3},\\
\frac{dz}{dt}&=v_z,&\frac{dv_z}{dt}&=-\frac{\mu z}{r^3}+\varepsilon.
\end{align}
The only equilibrium point for this system is for $v_x=v_y=v_z=x=y=0$ and $z=\sqrt{\mu / \varepsilon}$. That is, the
test particle is stationary on the positive $z$ axis at a distance from the origin such that the Newtonian attraction and the external
acceleration field counterbalance each other. We refer to this unstable critical point as the \emph{cartesian stationary equilibrium}.

Back in parabolic coordinates and fictitious time $\tau$, a first straightforward observation is that the cartesian stationary equilibrium
cannot be handled in this coordinate system, as it corresponds to a position in which the azimuthal angle $\phi$ is undefined.
Secondly, since $d\phi/d\tau$ is a monotonic
function according to \eqref{eq:phi_diffeq}, it follows that there cannot be a parabolic stationary equilibrium point, and that only the coordinates $\xi$ and $\eta$
can be in a stationary point. From the definition \eqref{eq:xi_inv} we can deduce how a trajectory
in which $\xi$ is constant is constrained to a circular paraboloid symmetric with respect to the $z$ axis and defined by the equation
\begin{equation}
z = \frac{\xi_0^4-x^2-y^2}{2\xi_0^2},
\end{equation}
resulting from the inversion of eq. \eqref{eq:xi_inv}.
Similarly, a trajectory with constant $\eta$ will be constrained to the paraboloid defined by
\begin{equation}
z = \frac{x^2+y^2-\eta_0^4}{2\eta_0^2}
\end{equation}
(via inversion of eq. \eqref{eq:eta_inv}).

It is then interesting to note how a trajectory in which both $\xi$ and $\eta$ are constant will be constrained to the intersection of two coaxial circular
paraboloids with opposite orientation. That is, the trajectory will follow a circle centred on the $z$ axis and parallel to the $xy$ plane.
Additionally, according to eqs. \eqref{eq:fic_time} and \eqref{eq:phi_diffeq}, such a circular trajectory will have constant angular velocity both in fictitious
and real time. Such orbits are known in the literature as \emph{static orbits} \citep{forward_statite_1991},
\emph{displaced circular orbits} \citep{dankowicz_special_1994,lantoine_complete_2011}, \emph{displaced non-Keplerian orbits}
\citep{mcinnes_dynamics_1998}, or \emph{sombrero orbits} \citep{namouni_accelerated_2007}.

From a physical point of view, displaced circular orbits are possible when the initial conditions satisfy the following requirements:
\begin{itemize}
 \item the distance from the $xy$ plane is such that the net force acting on the test particle is perpendicular to
 the $z$ axis (i.e., the total force has zero $z$ component),
 \item the initial velocity vector is lying on the plane $\Pi$ of the displaced circular orbit, it is perpendicular
 to the projection of the position vector on $\Pi$ and its magnitude has the same value it would assume in a circular Keplerian orbit
 with a fictitious central body lying in correspondence of the $z$ axis on the $\Pi$ plane
 (where the mass of the fictitious body is generating the total force experienced by the test particle).
\end{itemize}
In other words, with these initial conditions the test particle evolves along a Keplerian planar circular orbit under the influence of a fictitious body
lying on the positive $z$ axis. These requirements are satisfied by the following cartesian initial conditions:
\begin{align}
\boldsymbol{r}_0 & = \left(\sqrt{\left(\frac{z\mu}{\varepsilon} \right)^\frac{2}{3}-z^2},0,z \right),\label{eq:dco_init_cond}\\
\boldsymbol{v}_0 & = \left(0,\sqrt{\frac{\varepsilon}{z}\left[\left(\frac{z\mu}{\varepsilon} \right)^\frac{2}{3}-z^2\right]},0\right),\label{eq:dco_init_vel}
\end{align}
where $z>0$ and where we have taken advantage of the cylindrical symmetry of the problem by choosing,
without loss of generality, a set of initial conditions on the $xz$ plane. It is clear from eqs. \eqref{eq:dco_init_cond} and \eqref{eq:dco_init_vel} that there
exist a limit on the value of $z$ after which displaced circular orbits are not possible because the radicand in the expression
for the $x$ coordinate
becomes negative. Physically, this means that the gravitational force cannot counterbalance the constant acceleration field in the
$z$ direction. This limit value is clearly in correspondence of the cartesian stationary equilibrium.

From a mathematical point of view, a displaced circular orbit must turn the solutions $\xi\left(\tau\right)$ and $\eta\left(\tau\right)$ into constants.
From eqs. \eqref{eq:xi_simpl} and \eqref{eq:eta_simpl} it is clear that these expressions can become constants only when
$f_{\xi}^{\prime}\left(\frac{\xi_{r}^{2}}{2}\right)$ and $f_{\eta}^{\prime}\left(\frac{\eta_{r}^{2}}{2}\right)$ are zero.
This condition is equivalent to the requirement that the two polynomials $f_\xi$ and $f_\eta$ have roots of multiplicity greater than one.
From the point of view of the theory of dynamical systems, the two polynomials need to have roots of multiplicity greater than one because
otherwise the zeroes of the differential equations \eqref{eq:xi_diffeq} and \eqref{eq:eta_diffeq} are in correspondence of a point
in which the equations lose their properties of differentiability and Lipschitz continuity, and the resulting equilibria are thus spurious.

It can be verified by direct substitution that the initial conditions \eqref{eq:dco_init_cond} and \eqref{eq:dco_init_vel}, after the transformation into parabolic coordinates,
are roots of both the characteristic polynomials $f_\xi$ and $f_\eta$ and of their derivatives. Our solution in terms of Weierstrassian functions is
thus consistent with known results \citep[e.g., see][]{namouni_accelerated_2007} regarding the existence and characterisation of the equilibrium points in the Stark problem.
\section{Conclusions}
In this paper we introduced a new solution to the  Stark problem based on Weierstrass elliptic and related functions. Our treatment yields an exact (i.e., non-perturbative)
and explicit solution of the full three-dimensional problem in terms of a set of unique formul\ae{} valid for all initial conditions and physical parameters of the system.
Formally, the result is remarkably similar to the solution of the two-body problem: the evolution of the coordinates is given as a function of an anomaly (or, a fictitious time)
connected to the real time by a transcendental equation.

The simplicity of our formulation allows us to derive several new results. In particular, we were able to formulate conditions for the existence of quasi-periodic and periodic orbits,
and to successfully identify instances of (quasi) periodic orbits using numerical techniques. We were also able to formulate a new simple analytical criterion
to study the boundness of the motion, a result that can be particularly interesting for astrodynamical applications (e.g., in the study of the ejection of dust grains in the
outer Solar System -- see \citet{belyaev_dynamics_2010} and \cite{pastor_influence_2012}). Another result of astrodynamical interest (in connection
to the design of powered flyby manoeuvres) is the
identification of an analytical formula for the determination of the orientation of the asymptotic planes of motion at infinity in case of unbound orbits.

Our analysis shows how the Weierstrassian formalism can be fruitfully applied to yield a new insight in the dynamics
of the Stark problem. We hope that our results will contribute to revive the interest in this beautiful and powerful mathematical tool.
\section*{Acknowledgements}
F. Biscani would like to thank Dr. Santiago Nicolas Lopez Carranza for helpful discussion, and E. S.
for providing the motivation to complete the manuscript.

The authors would also like to thank the reviewers, Prof. Ryan Russell, Noble Hatten and Nick Bradley, for their insightful
input and suggestions during the review process.
\bibliographystyle{mn2e}
\bibliography{biblio}

\appendix

\section{Implementation details}
\subsection{Implementation of the Weierstrassian functions}
The Weierstrassian functions are not as readily available in scientific computation packages as other special functions.
Following \citet[\S 18.9 and \S 18.10]{abramowitz_handbook_1964}, it is possible to express them in terms of Jacobi elliptic and theta
functions.
The recipes in \citet{abramowitz_handbook_1964} do not present particular difficulties in terms of implementation details. A minor complication is that
the cases in which the Weierstrass invariant $g_3$ is negative are transformed in non-negative $g_3$ via the homogeneity relation
\begin{equation}
\wp\left(z;g_2,g_3\right) = -\wp\left(\imath z;g_2,-g_3\right)\label{eq:homo_g3}
\end{equation}
(and similar relations hold for $\zeta$ and $\sigma$). This transformation is not problematic for the computation of the values of the functions,
but it needs to be properly taken into account when computing auxiliary quantities such as the half-periods and the roots of the characteristic
cubic equations.

Regarding the half-periods, it is seen from eq. \eqref{eq:homo_g3} how the effect of the homogeneity relation is that of a rotation of the half-periods
by $-\pi/2$ in the complex plane (via the $\imath$ factor applied to the argument $z$ on the right-hand side). The half-periods can then be first calculated
in the transformed non-negative $g_3$ case, and afterwards they can be rotated back to obtain the original half-periods.

Regarding the roots of the characteristic polynomial
\begin{equation}
y = 4x^3 - g_2x - g_3,
\end{equation}
one can see how a change in sign in $g_3$ corresponds to a reflection with respect to both the $x$ and $y$ axes. The
net effect will thus be equivalent to a simple change of the sign of all roots.

For the actual implementation of the Weierstrassian functions, we used the elliptic functions module of the multiprecision Python library \emph{mpmath} \citep{mpmath}.
\subsection{On the computation of the complex logarithms in equation \eqref{eq:phi_tau}}
\label{sec:complex_log}
The solution for the evolution of the $\phi$ coordinate in fictitious time, eq. \eqref{eq:phi_tau}, involves, in the general case, the computation
of complex logarithms. Since the complex logarithm is a multivalued function, care must be taken in order to select values that yield physically meaningful solutions.

The standard way of proceeding when dealing with complex logarithms is to restrict the computation to the principal value $\Log$ of the logarithm,
i.e., the unique value whose imaginary
part lies in the interval $\left(-\pi,\pi\right]$. In doing so, if one takes the logarithm of a complex function whose values cross the negative real axis
(i.e., the branch cut of $\Log$),
a discontinuity will arise -- the imaginary part of the logarithm of the function will jump from $\pi$ to $-\pi$ (or vice versa).
In the case of the Stark problem, this means
that $\phi\left(\tau\right)$ will be discontinuous. These discontinuities are merely an artefact of the way of choosing a particular logarithm
value among all the possible ones, and they need to be dealt with in order to produce a physically correct (i.e., continuous) solution.

We start by recalling the following series expansion for the logarithm of
$\sigma$ \citep[][\S CVI]{jules_tannery_elements_1893}:
\begin{multline}
\Log\sigma\left(u\right)=\Log\frac{2\omega_{R}}{\pi}+\frac{\eta_Ru^{2}}{2\omega_{R}}+\Log\sin\frac{\pi u}{2\omega_{R}}\\
+\sum_{r=1}^{\infty}\frac{q^{2r}}{r\left(1-q^{2r}\right)}\left(2\sin\frac{r\pi u}{2\omega_{R}}\right)^{2},\label{eq:log_exp}
\end{multline}
where $\eta_{R}=\zeta\left(\omega_{R}\right)$ and $q=\exp\left(\imath\pi\frac{\omega_{C}}{\omega_{R}}\right)$,
and $u$ is decomposed into its components along the fundamental periods as
\begin{equation}
u=2\alpha\omega_{R}+2\beta\omega_{C},
\end{equation}
with $\alpha,\beta\in\mathbb{R}$. This series expansion is convergent
for $\left|\beta\right|<1$, or, in other words, as long as $u$ is
confined to the strip in the complex plane defined by $\left|\Im\left(u\right)\right|<2\Im\left(\omega_{C}\right)$.

We turn now to the study of the behaviour of the series expansion
\eqref{eq:log_exp} within the real period $2\omega_{R}$ and in the positive half
of the strip of convergence. That is, we study the behaviour of the
series expansion of $\Log\sigma\left[x_{\ast}+\imath2\beta\Im\left(\omega_{C}\right)\right]$,
with $x_{\ast}$ as a real variable in the interval $\left[0,2\omega_{R}\right)$
and $0<\beta<1$. We first note that, from eq. \eqref{eq:log_exp}, there exists a
potential discontinuity in the computation of the complex logarithm
\begin{equation}
\Log\sin\frac{\pi\left[x_{\ast}+\imath2\beta\Im\left(\omega_{C}\right)\right]}{2\omega_{R}},\label{eq:log_sin}
\end{equation}
when its argument crosses the negative real axis. However, by applying
elementary trigonometric identities, we can write
\begin{align}
\Re\left\{ \sin\frac{\pi\left[x_{\ast}+\imath2\beta\Im\left(\omega_{C}\right)\right]}{2\omega_{R}}\right\}  & =\sin\frac{\pi x_{\ast}}{2\omega_{R}}\cosh\frac{\pi\beta\Im\left(\omega_{C}\right)}{\omega_{R}},\label{eq:R_log}\\
\Im\left\{ \sin\frac{\pi\left[x_{\ast}+\imath2\beta\Im\left(\omega_{C}\right)\right]}{2\omega_{R}}\right\}  & =\cos\frac{\pi x_{\ast}}{2\omega_{R}}\sinh\frac{\pi\beta\Im\left(\omega_{C}\right)}{\omega_{R}}.
\end{align}
That is, the argument of the logarithm in \eqref{eq:log_sin} crosses the real axis
when $x_{\ast}=\omega_{R}$. But then, for $x_{\ast}=\omega_{R}$,
the real part \eqref{eq:R_log} of the argument of the logarithm is strictly positive
(as the hyperbolic cosine is a strictly positive function), and hence
the crossing of the real axis does not happen in correspondence of
the branch cut of the principal value of the logarithm. This means
that, for $x_{\ast}\in\left[0,2\omega_{R}\right)$, the series expansion
\eqref{eq:log_exp} of $\Log\sigma\left[x_{\ast}+\imath2\beta\Im\left(\omega_{C}\right)\right]$
is a continuous function.

Outside the interval $\left[0,2\omega_{R}\right)$, we can represent
a variable $x\in\mathbb{R}$ as $x=x_{\ast}+2N\omega_{R}$, where
$N\in\mathbb{Z}$. Recalling now the definition of the Weierstrass
sigma function \citep[][\S 195]{greenhill_applications_1959}, we can write
\begin{multline}
\sigma\left[x+\imath 2\beta\Im\left(\omega_{C}\right)\right]=\sigma\left[x_{\ast}+2N\omega_{R}+\imath2\beta\Im\left(\omega_{C}\right)\right]\\
=\exp\left\{ \Log\left[x_{\ast}+2N\omega_{R}+\imath 2\beta\Im\left(\omega_{C}\right)\right]+
\vphantom{\int_{0}^{x_{\ast}+2N\omega_{R}+\imath2\beta\Im\left(\omega_{C}\right)}\left[\zeta\left(z\right)-\frac{1}{z}\right]dz}
\right.\\
\left.\int_{0}^{x_{\ast}+2N\omega_{R}+\imath2\beta\Im\left(\omega_{C}\right)}\left[\zeta\left(z\right)-\frac{1}{z}\right]dz\right\}.\label{eq:sigma_def}
\end{multline}
We can split the integral in eq. \eqref{eq:sigma_def} as
\begin{multline}
\int_{0}^{x_{\ast}+2N\omega_{R}+\imath2\beta\Im\left(\omega_{C}\right)}\left[\zeta\left(z\right)-\frac{1}{z}\right]dz\\
=\int_{0}^{x_{\ast}+\imath2\beta\Im\left(\omega_{C}\right)}\left[\zeta\left(z\right)-\frac{1}{z}\right]dz\\
+\int_{x_{\ast}+\imath2\beta\Im\left(\omega_{C}\right)}^{x_{\ast}+2N\omega_{R}+\imath2\beta\Im\left(\omega_{C}\right)}
\left[\zeta\left(z\right)-\frac{1}{z}\right]dz,
\label{eq:three_integrals}
\end{multline}
and, following \citep[][\S CXVII]{jules_tannery_elements_1893}, the third integral in eq. \eqref{eq:three_integrals} can be computed as
\begin{multline}
\int_{x_{\ast}+\imath2\beta\Im\left(\omega_{C}\right)}^{x_{\ast}+2N\omega_{R}+\imath2\beta\Im\left(\omega_{C}\right)}\left[\zeta\left(z\right)-\frac{1}{z}\right]dz\\
=\Log\left[x_{\ast}+\imath2\beta\Im\left(\omega_{C}\right)\right]-\Log\left[x_{\ast}+2N\omega_{R}+\imath2\beta\Im\left(\omega_{C}\right)\right]\\
+2N\eta_{R}\left[x_{\ast}+\imath2\beta\Im\left(\omega_{C}\right)+N\omega_{R}\right]-\imath N\pi.
\end{multline}
In other words,
\begin{multline}
\Log\sigma\left[x+\imath2\beta\Im\left(\omega_{C}\right)\right]=\Log\sigma\left[x_{\ast}+\imath2\beta\Im\left(\omega_{C}\right)\right]\\
+2N\eta_{R}\left[x_{\ast}+\imath2\beta\Im\left(\omega_{C}\right)+N\omega_{R}\right]-\imath N\pi,\label{eq:final_log_sigma}
\end{multline}
which corresponds to the homogeneity relation in \citet[][\S 18.2]{abramowitz_handbook_1964}. Since, as we have
seen, $\Log\sigma\left[x_{\ast}+\imath2\beta\Im\left(\omega_{C}\right)\right]$
is a continuous function, the only possible discontinuities in eq. \eqref{eq:final_log_sigma}
are in the neighbourhood of $x=2N\omega_{R}$, where $x_{\ast}$
changes discontinuously by $\pm2\omega_{R}$ and $N$ by $\pm1$. For
$x=2N\omega_{R}$, $x_{\ast}$ is zero and the limit from the right
is
\begin{multline}
L^{+}=\lim_{x\to\left(2N\omega_{R}\right)^{+}}\Log\sigma\left[x+\imath2\beta\Im\left(\omega_{C}\right)\right]\\
=\Log\sigma\left[\imath2\beta\Im\left(\omega_{C}\right)\right]+2N\eta_{R}\left[\imath2\beta\Im\left(\omega_{C}\right)+N\omega_{R}\right]-\imath N\pi.
\end{multline}
The limit from the left instead is
\begin{multline}
L^{-}=\lim_{x\to\left(2N\omega_{R}\right)^{-}}\Log\sigma\left[x+\imath2\beta\Im\left(\omega_{C}\right)\right]\\
=\Log\sigma\left[2\omega_{R}+\imath2\beta\Im\left(\omega_{C}\right)\right]\\
+2\left(N-1\right)\eta_{R}\left[\imath2\beta\Im\left(\omega_{C}\right)+\left(N+1\right)\omega_{R}\right]-\imath\left(N-1\right)\pi.
\end{multline}
By using the series expansion \eqref{eq:log_exp}, we can write
\begin{multline}
L^{+}=\Log\frac{2\omega_{R}}{\pi}+\frac{\eta_{R}\left[\imath2\beta\Im\left(\omega_{C}\right)\right]^{2}}{2\omega_{R}}+
\Log\sin\frac{\imath\pi\beta\Im\left(\omega_{C}\right)}{\omega_{R}}\\
+\sum_{r=1}^{\infty}\frac{q^{2r}}{r\left(1-q^{2r}\right)}\left\{ 2\sin\frac{r\pi\left[\imath2\beta\Im\left(\omega_{C}\right)\right]}{2\omega_{R}}\right\} ^{2}\\
+2N\eta_{R}\left[\imath2\beta\Im\left(\omega_{C}\right)+N\omega_{R}\right]-\imath N\pi
\end{multline}
and
\begin{multline}
L^{-}=\Log\frac{2\omega_{R}}{\pi}+\frac{\eta_{R}\left[2\omega_{R}+\imath2\beta\Im\left(\omega_{C}\right)\right]^{2}}{2\omega_{R}}\\
+\Log\sin\frac{\pi\left[2\omega_{R}+\imath2\beta\Im\left(\omega_{C}\right)\right]}{2\omega_{R}}\\
+\sum_{r=1}^{\infty}\frac{q^{2r}}{r\left(1-q^{2r}\right)}\left\{ 2\sin\frac{r\pi\left[2\omega_{R}+\imath2\beta\Im\left(\omega_{C}\right)\right]}
{2\omega_{R}}\right\} ^{2}\\
+2\left(N-1\right)\eta_{R}\left[\imath2\beta\Im\left(\omega_{C}\right)+\left(N+1\right)\omega_{R}\right]-\imath\left(N-1\right)\pi.
\end{multline}
By noting that
\begin{equation}
\Log\left[\pm\sin\frac{\imath\pi\beta\Im\left(\omega_{C}\right)}{\omega_{R}}\right]=
\Log\sinh\frac{\pi\beta\Im\left(\omega_{C}\right)}{\omega_{R}}\pm\imath\frac{\pi}{2}
\end{equation}
(as $\beta$, $\Im\left(\omega_{C}\right)$ and $\omega_{R}$ are
all real positive quantities by definition), it can be verified, after
a few algebraic passages, that $L^{+}=L^{-}$, and thus the right-hand side of eq. \eqref{eq:final_log_sigma} is a continuous
function.

Going back to the Stark problem, we can immediately see how the logarithmic forms in eq. \eqref{eq:phi_tau} are in the same form as in
eq. \eqref{eq:final_log_sigma}. For instance, in
\begin{equation}
\ln\sigma_\xi\left(\tau-\tau_\xi+u_\xi\right)
\end{equation}
the real variable is $\tau$, while $u_\xi$ is defined as
\begin{equation}
u_\xi=\wp_\xi^{-1}\left(-\frac{\delta_\xi}{\gamma_\xi}\right).
\end{equation}
Since $u_\xi$ is the result of an inverse $\wp$, it can always be chosen inside the fundamental period parallelogram, where the condition
of convergence of the series expansion \eqref{eq:log_exp} ($\left| \beta \right| < 1$) is always satisfied\footnote{From 
the point of view of practical implementation, one can choose among two possible values for $u_\xi$ in the fundamental period parallelogram.
In order to improve the convergence properties of the series expansion, it is convenient to select the value with the smaller imaginary part.}.
Eq. \eqref{eq:final_log_sigma} can thus be substituted into eq. \eqref{eq:phi_tau} to provide a formula for $\phi\left(\tau\right)$ free
of discontinuities.
\subsection{Solution algorithm}
\label{subsec:sol_algorithm}
In this section, we are going to detail the steps of a possible implementation of our solution to the Stark problem, starting from
initial conditions in cartesian coordinates. The
algorithm outlined below requires the availability of an implementation of the Weierstrassian functions $\wp$, $\wp^\prime$, $\wp^{-1}$,
$\zeta$ and $\sigma$, and of a few related ancillary functions (e.g., for the conversion of the invariants $g_2$ and $g_3$ to the half-periods
$\omega$ and $\omega^\prime$). Chapter 18 in \citet{abramowitz_handbook_1964} details how to implement these requirements in terms of Jacobi
theta and elliptic functions and integrals.

The algorithm is given as follows:
\begin{enumerate}
\item transform the initial cartesian coordinates into parabolic coordinates via eqs. \eqref{eq:xi_inv}-\eqref{eq:phi_inv},
and compute the initial Hamiltonian momenta $p_\xi$, $p_\eta$ and $p_\phi$ via eqs. \eqref{eq:p_xi}-\eqref{eq:p_phi};
\item compute the constants of motion $h$, $\alpha_1$ and $\alpha_2$, through the substitution of the initial Hamiltonian
coordinates and momenta into eqs. \eqref{eq:Ham_def}, \eqref{eq:alpha1} and \eqref{eq:alpha2};
\item calculate the roots of the bicubic polynomials on the right-hand sides of eqs. \eqref{eq:xi_diffeq} and \eqref{eq:eta_diffeq}.
Among the positive roots,
choose one for each of the two polynomials as the $\xi_r$ and $\eta_r$ values. In the case of the $\xi$
coordinate, $\xi_r$ must be a \emph{reachable} root, i.e., a value that will actually be assumed by $\xi$
at some point in time\footnote{Consider, for instance, a phase space portrait like the one depicted in Figure \ref{fig:tridimensional_case}(b).
Depending on the initial conditions, the test particle will be confined either to a circulation lobe (in which case there are
two reachable roots, where the lobe intersects the horizontal axis) or to the parabolic arm (in which case there is only
one reachable root, where the parabolic arm intersects the horizontal axis).};
\item compute the fictitious times of ``pericentre passage'' $\tau_\xi$ and $\tau_\eta$ via eq. \eqref{eq:depressed_transform}. The integral
can be solved either via the inverse Weierstrass function \citep{hoggatt_inverse_1955} or via elliptic integrals
\citep[e.g.,][\S 3.131 and \S 3.138]{gradshtein_table_2007}. The signs of $\tau_\xi$ and $\tau_\eta$ must be chosen
in accordance with the choice of $\xi_r$ and $\eta_r$ and with the initial signs of $p_\xi$ and $p_\eta$. For instance,
in our implementation of this algorithm we always pick as $\xi_r$ the smallest reachable root, so that the sign of
$\tau_\xi$ is the opposite of the sign of the initial value of $p_\xi$ (i.e., if initially $p_\xi < 0$ then $\xi_r$ will be reached
in the future and thus $\tau_\xi > 0$);
\item at this point, it will be possible to compute the evolution in fictitious time of $\xi$, $\eta$ and $\phi$ via eqs.
\eqref{eq:xi_simpl}, \eqref{eq:eta_simpl} and \eqref{eq:phi_tau}.
The complex logarithm appearing in the equation for $\phi$, eq. \eqref{eq:phi_tau}, should be computed using the methodology described in
Appendix \ref{sec:complex_log} in order to avoid discontinuities;
\item in order to compute the time equation, eq. \eqref{eq:time_equation}, determine the roots $e_i$ of the characteristic cubic equations
\eqref{eq:char_cubic} and the fundamental half-periods $\omega_i$ they correspond to, as explained in \S \ref{sec:time_equation}. It will
now be possible to compute $t\left( \tau \right)$, and to invert it via numerical techniques to yield $\tau\left( t \right)$.

\end{enumerate}
\section{Code availability}
The Weierstrassian functions, the analytical formul\ae{} presented in this paper,
and the algorithm outlined in Appendix \ref{subsec:sol_algorithm}
have been implemented in the Python programming language.
The implementation is available under an open-source license from the code repository
\newline
\newline
\url{https://github.com/bluescarni/stark_weierstrass}
\newline
\newline
\bsp

\label{lastpage}

\end{document}